\documentclass{amsproc}

\usepackage{epsfig}

\theoremstyle{definition}

\theoremstyle{remark}

\numberwithin{equation}{section}

\newcommand{\C}{\mathbb C}
\newcommand{\R}{\mathbb R}
\newcommand{\Z}{\mathbb Z}

\newcommand{\T}{\mathbb T}

\newcommand{\m}{\omega}

\newcommand{\be}{\begin{equation}}
\newcommand{\ee}{\end{equation}}
\newcommand{\lab}[1]{\label{#1}}
\def\stackreb#1#2{\ \mathrel{\mathop{#1}\limits_{#2}}}

\begin{document}

\title[Elliptic beta integrals and solvable models of statistical mechanics]
{Elliptic beta integrals and solvable \\ models of statistical mechanics}

\author{V. P. Spiridonov}
\address{Max-Planck-Institut f\"ur Mathematik,
 Vivatsgasse 7, 53111, Bonn, Germany}
\curraddr{Bogoliubov Laboratory of Theoretical Physics, JINR,
 Dubna, Moscow reg. 141980, Russia}
\email{spiridon@theor.jinr.ru}
\thanks{Work was supported in part by the Russian foundation for basic research
(RFBR grant no. 09-01-00271).\\
\indent Date: 30 May 2010. \\
\indent
To appear in the Proceedings of the Jairo Charris Seminar 2010
``Algebraic Aspects of Darboux Transformations,
Quantum Integrable Systems and Supersymmetric Quantum Mechanics",
Contemp. Math., Amer. Math. Soc., Providence, RI.
}



\begin{abstract}
The univariate elliptic beta integral was discovered by the author in 2000.
Recently Bazhanov and Sergeev have interpreted it as a star-triangle
relation (STR). This important observation is discussed in more detail
in connection to author's previous work on the elliptic modular double
and supersymmetric dualities.
We describe also a new Faddeev-Volkov type solution of STR, connections
with the star-star relation, and higher-dimensional analogues of such
relations. In this picture, Seiberg dualities are described by symmetries
of the  elliptic hypergeometric integrals (interpreted as superconformal
indices) which, in turn, represent STR and Kramers-Wannier type duality
transformations for elementary partition functions
in solvable models of statistical mechanics.
\end{abstract}

\maketitle
\tableofcontents

\section{The simplest elliptic hypergeometric integrals}

In the present paper we discuss relations between a new class
of special functions, called elliptic hypergeometric functions,
and solvable models of statistical mechanics. We describe the most
complicated known integrable systems defined on $2d$ (two-dimensional)
lattices representing continuous spin generalizations of the
well known Ising model and its various extensions. Actually,
these novel integrable models correspond to some discretized $2d$
quantum field theories.
Also we indicate connections with the  $4d$
supersymmetric field theories, where elliptic hypergeometric
integrals have found recently the major application.
We start from a brief technical introduction to the needed results on
special functions and discuss
the physical systems they apply to in the following chapters.

General theory of elliptic hypergeometric integrals was
formulated in \cite{spi:umn,spi:theta,spi:thesis}.
We skip the structural definition of these integrals and refer for the corresponding
details to a reasonably short survey given in \cite{spi:essays}.

Let us denote
$$
(z;q)_\infty=\prod_{k=0}^\infty(1-zq^k),\ \ \ |q|<1, \quad z\in\C,
$$
the standard infinite $q$-product and
$$
\Gamma(z;p,q)= \prod_{i,j=0}^\infty
\frac{1-z^{-1}p^{i+1}q^{j+1}}{1-zp^iq^j}, \quad |p|, |q|<1,\quad z\in\C^*,
$$
the standard elliptic gamma function. Below we use the conventions
\begin{eqnarray*}
&& \Gamma(a,b;p,q):=\Gamma(a;p,q)\Gamma(b;p,q), \ \ \
\Gamma(az^{\pm1};p,q):=\Gamma(az;p,q)\Gamma(az^{-1};p,q),
\\ &&
\Gamma(az^{\pm1}y^{\pm1};p,q):=\Gamma(azy;p,q)\Gamma(az^{-1}y;p,q)
\Gamma(azy^{-1};p,q)\Gamma(az^{-1}y^{-1};p,q).
\end{eqnarray*}

One has the symmetry
$\Gamma(z;p,q)= \Gamma(z;q,p)$ and the inversion formula
$$
\Gamma\Big(a,\frac{pq}{a};p,q\Big)=1,\ \ \  \text{or} \  \
\Gamma(a,a^{-1};p,q)=\frac{1}{\theta(a;p)\theta(a^{-1};q)},
$$
which follows from the difference equations
$$
\Gamma(qz;p,q)=\theta(z;p)\Gamma(z;p,q), \qquad
\Gamma(pz;p,q)=\theta(z;q)\Gamma(z;p,q),
$$
where
$$
\theta(z;p)=(z;p)_\infty (pz^{-1};p)_\infty
$$
is a theta function. The standard odd Jacobi theta
function has the form \cite{whi-wat}
\begin{eqnarray}\nonumber &&
 \theta_1(u|\tau)=-\theta_{11}(u)=-\sum_{k\in \Z}e^{\pi \textup{i} \tau(k+1/2)^2}
e^{2\pi \textup{i} (k+1/2)(u+1/2)}
\\  && \makebox[8em]{}
=\textup{i}p^{1/8}e^{-\pi \textup{i} u}(p;p)_\infty\theta(e^{2\pi \textup{i} u};p),
\nonumber\end{eqnarray}
where we denoted $p=e^{2\pi \textup{i} \tau}$.

The univariate elliptic beta integral \cite{spi:umn} forms a
cornerstone of a new powerful class of exactly computable integrals.
It is described by the following explicit formula
\begin{equation}
\kappa \int_{\mathbb T}\frac{\prod_{j=1}^6
\Gamma(t_jz^{{\pm 1}};p,q)}{\Gamma(z^{\pm 2};p,q)}\frac{dz}{\textup{i} z}
= \prod_{1\leq j<k\leq6}\Gamma(t_jt_k;p,q),
\label{ell-int}\end{equation}
where $\mathbb T$ is the unit circle with positive orientation,
$$
\kappa= \frac{(p;p)_\infty (q;q)_\infty}{4\pi},
$$
and six complex parameters $t_j,\; j=1,\ldots,6$,
satisfy the inequalities $|t_j|<1$ and the balancing condition
\begin{equation}
\prod_{j=1}^6 t_j=pq.
\label{balancing}\end{equation}

We use the word ``integral" in two meanings. When referred to the
exactly computable cases, like \eqref{ell-int} or the standard Euler beta integral
lying on its bottom, it means either the function defined by the left-hand
side or, more often, the whole identity. In other cases it means an integral
representation for a function of interest or a class of functions with
common structure.

As shown in \cite{spi:theta}, the left-hand side of relation (\ref{ell-int})
serves as the orthogonality measure for the most general known family of
biorthogonal functions with the properties characteristic to
classical orthogonal polynomials (Chebyshev, Hermite, Laguerre, Jacobi, $\ldots$ ,
Askey-Wilson polynomials). In the same paper the elliptic beta
integral has been generalized to the following function
\begin{equation}
V(t_1,\ldots,t_8;p,q)=\kappa
\int_{\mathbb T}\frac{\prod_{j=1}^8
\Gamma(t_jz^{{\pm 1}};p,q)}{\Gamma(z^{\pm 2};p,q)}\frac{dz}{\textup{i} z},
\label{eghf}\end{equation}
where $|t_j|<1$ and $\prod_{j=1}^8t_j=(pq)^2.$ This is a natural elliptic analogue
of the Gauss hypergeometric function since its features generalize most of the
special function properties of the $_2F_1$-series \cite{spi:thesis,spi:essays}.
For $t_jt_k=pq$, $j\neq k$, $V$-function reduces to the elliptic beta integral
and, for this reason, it can be called the elliptic beta integral of a higher order.

In \cite{spi:bai}, the author has introduced the
following universal  integral transformation for functions analytical in
the vicinity of the  unit circle $\T$:
\begin{equation}
g(w;t)=\kappa\int_{\T}\Delta(t;w,z;p,q)f(z;t)
\frac{dz}{\textup{i} z},
\label{trn} \end{equation}
where the kernel
\begin{equation}
\Delta(t;w,z;p,q):=\Delta(t;w,z)=\Gamma(tw^{\pm1}z^{\pm1};p,q),
\ \ |t|<1,
\label{ker} \end{equation}
is a particular product of four elliptic gamma functions.
In \cite{spi-war:inversions}, it was shown that this integral transformation
obeys the key property making it very similar to the Fourier
transformation. Namely, its inverse is obtained essentially by
the reflection $t\to t^{-1}$.

An explicit example of the pair of functions
$g(w;t)$ and $f(z;t)$ in \eqref{trn} can be easily
found from the elliptic beta integral.
Indeed, let us denote $t_5=tw$ and $t_6=tw^{-1}$ (so that
$t^2\prod_{j=1}^4 t_j=pq$). Then,
\begin{eqnarray}\label{a_1}
&& f(z;t)=\frac{\prod_{j=1}^4\Gamma(t_jz^{\pm1};p,q)}
{\Gamma(z^{\pm 2};p,q)},
\\ &&
g(w;t)=\Gamma(t^2;p,q)\prod_{1\leq i<j\leq 4}\Gamma(t_it_j;p,q)
\prod_{j=1}^4\Gamma(tt_jw^{\pm1};p,q),
\label{bp}\end{eqnarray}
where $|tw^{\pm1}|, |t_j|<1$.

Because of the permutational symmetry, any of the original variables $t_j$
can be associated with the distinguished parameter $t$.
After fixing $t_1=sy, t_2=sy^{-1}$ and $t_3=rx, t_4=rx^{-1}$,
one can rewrite the elliptic beta integral in the form
\begin{eqnarray}\nonumber
&& \int_{\T}\varphi(z)\Delta(r;x,z)\Delta(s;y,z)\Delta(t;w,z)\frac{dz}{ \textup{i}  z}
\\ && \makebox[2em]{}
=\chi(r,s,t)\Delta(rs;x,y)\Delta(rt;x,w)\Delta(st;y,w),
\label{mstr} \end{eqnarray}
where $rst=\pm \sqrt{pq}$ and
\begin{eqnarray}\nonumber
&& \varphi(z)= \frac{(p;p)_\infty (q;q)_\infty}{4\pi\Gamma(z^{\pm2};p,q)}
=\frac{1}{4\pi}(p;p)_\infty (q;q)_\infty\theta(z^2;p)\theta(z^{-2};q),
\\ \makebox[0em]{}
&& \chi(r,s,t)=\Gamma(r^2,s^2,t^2;p,q).
\label{norm} \end{eqnarray}

A key application of definition \eqref{trn} consists in the construction of
a tree of identities for multiple elliptic hypergeometric integrals with
many parameters \cite{spi:bai}. Using one of the corresponding symmetry transformations,
the following relation has been derived in \cite{spi:con}
\be
\phi(x;c,d|\xi;s)=\kappa\int_\T R(c,d,a,b;x,w|s)\phi(w;a,b|\xi;s)\frac{dw}{\textup{i} w},
\lab{key-rel}\ee
where the ``basis vector" $\phi$ has the form
\be
\phi(w;a,b|\xi;s)=\Gamma(sa\xi^{\pm1},sb\xi^{\pm1},
\sqrt{\frac{pq}{ab}}w^{\pm1}\xi^{\pm1}; p,q),
\lab{phi}\end{equation}
and the ``rotation" integral operator kernel is
\begin{eqnarray*}\nonumber
&& R(c,d,a,b;x,w|s)=
\frac{1}{\Gamma(\frac{pq}{ab},\frac{ab}{pq},w^{\pm2}; p,q)}
\\ && \makebox[4em]{} \times
V\left(s c,s d,\sqrt{\frac{pq}{cd}}x,
\sqrt{\frac{pq}{cd}}x^{-1},\frac{pq}{as},\frac{pq}{bs},
\sqrt{\frac{ab}{pq}}w,\sqrt{\frac{ab}{pq}}w^{-1};p,q\right).
\end{eqnarray*}
The function $\phi$ is a generalization of the kernel $\Delta(t;x,z)$,
since for $ab=pq/s^2$ one has the reduction
$$
\phi(w;a,\frac{pq}{as^2}|\xi;s)=\Delta(s;w,\xi).
$$
Using the $\Delta$-kernel, relation (\ref{key-rel}) was rewritten also
in \cite{spi:con} in a more compact form
\begin{eqnarray}\nonumber
&& \makebox[-1em]{}
\Delta(\alpha; x,\xi)\Delta(\beta;y,\xi)
=\kappa\int_\T r(\alpha,\beta,\gamma,\delta;x,y;t,w)
\Delta(\gamma; t,\xi)\Delta(\delta;w,\xi)\frac{dw}{\textup{i} w},
\\ &&  \makebox[-1em]{}
r(\alpha,\beta,\gamma,\delta;x,y;t,w)=\frac{1}
{\Gamma(\delta^{\pm2},w^{\pm2};p,q)}
V\left(\alpha x^{\pm1},\beta y^{\pm1},
\frac{pq}{\gamma} t^{\pm1},\frac{w^{\pm1}}{\delta}\right),
\nonumber\end{eqnarray}
where $\alpha\beta=\gamma\delta$ and
$$
V\left(\alpha x^{\pm1},\beta y^{\pm1},
\frac{pq}{\gamma} t^{\pm1},\frac{w^{\pm1}}{\delta}\right)
=\kappa\int_{\mathbb T}\frac{\Delta(\alpha;x,z)\Delta(\beta;y,z)
\Delta(\frac{pq}{\gamma};t,z)\Delta(\frac{1}{\delta};w,z)}
{\Gamma(z^{\pm 2};p,q)}\frac{dz}{\textup{i} z}.
$$
Here we use the condensed notation for parameters of the $V$-function:
$V(\ldots \alpha x^{\pm1}$ $\ldots)=V(\ldots \alpha x, \alpha x^{-1} \ldots)$.

The function $\phi$ emerges also in the context of the Sklyanin algebra \cite{skl}
(the algebra of the Yang-Baxter equation solutions),
\begin{eqnarray}\nonumber
&& S_\alpha S_\beta -S_\beta S_\alpha =\textup{i} (S_0S_\gamma+S_\gamma S_0),
\\ &&
S_0S_\alpha - S_\alpha S_0 = \textup{i} \frac{J_\beta-J_\alpha}{J_\gamma}
(S_\beta S_\gamma+S_\gamma S_\beta),
\label{s-rel2}\end{eqnarray}
where $J_{\alpha}$  are the structure constants and
$(\alpha,\beta,\gamma)$ is any cyclic permutation of $(1,2,3)$.
Namely, one has to consider the generalized
eigenvalue problems $A\phi=\lambda B\phi$, where $A$ and $B$ are linear combinations
of four generators $S_a,\ a=0,1,2,3,$ and $\lambda$ is a spectral parameter.
The function $\phi$ is defined uniquely up to multiplication by a constant
with the help of two such equations using a pair of Sklyanin algebras
forming an elliptic modular double \cite{spi:con}.  This algebra represents an
elliptic extension of the Faddeev modular double \cite{fad:mod}, but
there are actually two different modular doubles
at the elliptic level which obey different sets of involutions.

Relevance of the
Sklyanin algebra in this setting was noticed first by Rains \cite{rai:skl}.
For special quantized values of the parameters, the $\phi$-function
reduces to the intertwining vectors of Takebe \cite{tak},
which were used by Rosengren in \cite{ros} for the derivation of a discrete spin
version of relation \eqref{key-rel}.
In our case both Casimir operators of the algebra \eqref{s-rel2},
$K_0=\sum_{a=0}^3 S_a^2$ and $K_2=\sum_{\alpha=1}^3 J_\alpha S_\alpha^2$,
take continuous values, i.e. we deal with the continuous spin
representations related to the integral operator form
of the Yang-Baxter equation.

The following scalar product has been introduced in \cite{spi:con}
\begin{equation}
\langle \chi,\psi\rangle =\kappa\int_{\T}\frac{\chi(z)\psi(z)}
{\Gamma(z^{\pm 2};p,q)}\frac{dz}{\textup{i} z}.
\label{sp}\end{equation}
It has been shown that both the $V$-function itself and the
$\phi$-vectors form biorthogonal systems of functions with respect to this measure.
In particular, one has the relation
\begin{eqnarray}\nonumber
&& \kappa\int_\T \frac{\phi(e^{\textup{i} \varphi'};\frac{pq}{c},\frac{pq}{d}|\xi;s^{-1})
\phi(e^{\textup{i} \varphi};c,d|\xi;s)}
{\Gamma(\xi^{\pm2};p,q)}\frac{d\xi}{\textup{i} \xi}
\\ && \makebox[4em]{}
=\frac{2\pi}{(p;p)_\infty(q;q)_\infty}
\Gamma\left(\frac{pq}{cd},\frac{cd}{pq},e^{\pm2 \textup{i} \varphi};p,q
\right)\sqrt{1-v^2}\, \delta(v-v'),
\label{phi-bio}\end{eqnarray}
where $v=\cos\varphi$, $v'=\cos\varphi'$, and $\delta(v)$ is the Dirac delta-function.
(There is a missprint in formula (3.2) of \cite{spi:con} which misses the first
factor standing on the right-hand side of (\ref{phi-bio}).)
Positivity of the biorthogonality measure and of the $\phi$-function
corresponds to the unitarity of representations of the elliptic modular double.
Setting $cd=pq/s^2$, we obtain
\begin{eqnarray}\nonumber
&&
 (2\kappa)^2\int_{-1}^1 \frac{\Delta(s^{-1};e^{\textup{i} \varphi'},e^{\textup{i} \chi})
\Delta(s;e^{\textup{i} \varphi},e^{\textup{i} \chi})}
{\Gamma(e^{\pm 2\textup{i} \chi};p,q)}\frac{dX}{\sqrt{1-X^2}}
\\ && \makebox[4em]{}
=\Gamma\left(s^2,s^{-2},e^{\pm 2\textup{i} \varphi};p,q\right)\sqrt{1-v^2}\, \delta(v-v'),
\label{bio-weight}\end{eqnarray}
where $X=\cos\chi$.

The tetrahedral symmetry transformation for $V$-function, discovered in
\cite{spi:theta}, can be rewritten in the following form:
\begin{eqnarray}\nonumber
&& \makebox[-2em]{}
V(\alpha x^{\pm1}, \beta y^{\pm1}, \gamma w^{\pm1}, \delta z^{\pm1})
= \Gamma(\alpha^2,\beta^2,\gamma^2,\delta^2;p,q)\Delta(\alpha\beta;x,y)
\Delta(\gamma\delta;w,z)
\label{t1} \\ && \makebox[4em]{} \times
V(\sqrt{pq}\beta^{-1}x^{\pm1}, \sqrt{pq}\alpha^{-1}y^{\pm1},
\sqrt{pq}\delta^{-1}w^{\pm1}, \sqrt{pq}\gamma^{-1} z^{\pm1})
\\  && \makebox[2em]{}
=\Delta(\alpha\gamma;x,w)\Delta(\alpha\delta;x,z)\Delta(\beta\gamma;y,w)
\Delta(\beta\delta;y,z)
\nonumber \\ && \makebox[6em]{} \times
V(\beta x^{\pm1},\alpha y^{\pm1},\delta w^{\pm1},\gamma z^{\pm1})
\label{t2} \\ && \makebox[2em]{}
=\Gamma(\alpha^2,\beta^2,\gamma^2,\delta^2;p,q)\Delta(\alpha\beta;x,y)
\Delta(\alpha\gamma;x,w)\Delta(\alpha\delta;x,z)\Delta(\beta\gamma;y,w)
\label{t3} \\ && \makebox[-2em]{}\times
\Delta(\beta\delta;y,z)\Delta(\gamma\delta;w,z)
V(\sqrt{pq}\alpha^{-1}x^{\pm1}, \sqrt{pq}\beta^{-1}y^{\pm1},
\sqrt{pq}\gamma^{-1}w^{\pm1}, \sqrt{pq}\delta^{-1} z^{\pm1}),
\nonumber\end{eqnarray}
where $\alpha\beta\gamma\delta=\pm pq$. The latter two transformations are obtained
by repeated application of the first relation in combination with
permutation of the parameters. The full symmetry group of the $V$-function is
the Weyl group $W(E_7)$ for the exceptional root system $E_7$ \cite{Rains}. Therefore,
there are $72=\dim W(E_7)/S_8$ relations similar to \eqref{t1}, \eqref{t2},
\eqref{t3}, we just picked up three of them by breaking the $S_8$ permutational
symmetry and gathering the elliptic gamma functions into the $\Delta$-blocks.

The outstanding physical application of the elliptic beta integral has been
discovered by Dolan and Osborn \cite{DO}. They have shown that
the simplest superconformal (topological) indices of ${\mathcal N}=1$ supersymmetric
field theories coincide with known elliptic hypergeometric integrals.
Exact computability or the Weyl group symmetry transformations of
such integrals describe the Seiberg duality of ${\mathcal N}=1$
theories \cite{Seiberg}, since they prove coincidence of the
corresponding superconformal indices.

In this picture, the left-hand side of the univariate elliptic beta integral
evaluation formula \eqref{ell-int} describes the superconformal index
of the supersymmetric quantum chromodynamics with $SU(2)$ gauge group and $SU(6)$ flavor
group. This theory has one vector superfield (gauge fields) in the adjoint
representation of $SU(2)$ and a set of chiral superfields (matter fields)
in the fundamental representation of $SU(2)\times SU(6)$. The elementary particles
representing these fields describe the spectrum of the theory in the high energy limit,
where the coupling constant is vanishing due to the asymptotic freedom.
In the deep infrared region the theory is strongly coupled, all colored particles
confine, and one has the Wess-Zumino type model for mesonic fields lying in
the 15-dimensional totally antisymmetric tensor representation of $SU(6)$.
The superconformal index of the latter theory is described by the right-hand side
expression of formula \eqref{ell-int}. This construction gives a group-theoretical
interpretation of the elliptic beta integral.
After renormalizing the parameters $t_k=(pq)^{1/6}y_k$, the balancing condition
takes the form $\prod_{k=1}^6y_k=1$, which is nothing else than the
unitarity condition for the maximal torus variables of the group $SU(6)$.
This is the simplest example of
the Seiberg duality discovered in \cite{Seiberg}.  Further detailed investigation
of such interrelations and their consequences can be found in \cite{SV},
where many new elliptic beta integrals on root systems have been conjectured
and many new supersymmetric dualities have been found.

 The elliptic
hypergeometric integrals emerge also in the context of the
relativistic Calogero-Sutherland type models \cite{spi:cs}.
However, the first non-trivial example of the elliptic hypergeometric functions
was found from the exactly solvable models of statistical
mechanics. Namely, in \cite{fre-tur} Frenkel and Turaev have
shown that the Boltzmann weights (elliptic $6j$-symbols)
of the RSOS models of Date et al
\cite{djkmo:exactly}, generalizing Baxter's eight-vertex model
\cite{bax:partition},  are determined by particular
values of the terminating $_{12}V_{11}$ elliptic hypergeometric series
(in modern notations of \cite{spi:essays}). The same series has been
found by Zhedanov  and the author \cite{sz:cmp} in a completely different
setting, as a particular solution of the Lax pair equations for
a classical discrete integrable system.
In \cite{spi:special,spi:theta}, a family of meromorphic functions
obeying a novel two-index biorthogonality relation has been discovered.
It was explicitly conjectured in \cite{spi:special} that these functions
determine a new family
of solutions of the Yang-Baxter equation for discrete spin models.
Since the $V(t_1,\ldots,t_8;p,q)$
function is an integral generalization of the latter functions, in  \cite{spi:talk}
it was conjectured that the $V$-function determines a solution of
the Yang-Baxter equation.
A simple connection of the terminating $_{12}V_{11}$-series and
$V$-function with the Yang-Baxter equation for RSOS models
was discussed in \cite{KS}.
Recently, Bazhanov and Sergeev \cite{BS} have shown that
the elliptic beta integral can be rewritten as a star-triangle relation (STR)
which yields a new two-dimensional solvable model of statistical mechanics.
This is a new important application of integral \eqref{ell-int}
which is described in the next section. In this paper we show that the
symmetry transformations for the $V$-function have similar
interpretation as the star-star relations. Moreover, we conjecture
that all known exact formulas for elliptic hypergeometric integrals
describing the Seiberg duality transformations (at the level of
superconformal indices) \cite{SV}, in turn, represent STR and
Kramers-Wannier type duality transformations \cite{KW,W} for elementary partition
functions in solvable models of statistical mechanics \cite{bax:book}.

\section{The elliptic beta integral STR solution and star-star relation}

In \cite{BS}, Bazhanov and Sergeev have interpreted the elliptic
beta integral evaluation formula as a star-triangle relation
which gave a new solution of this relation. In order to describe it,
let us introduce the parameter $\eta$ related to the bases $p$ and $q$ as
$$
e^{-2\eta}=pq
$$
and pass to the additive notation
$$
z=e^{ \textup{i}  u},\quad
x\to e^{ \textup{i}  x},\quad
y\to e^{ \textup{i}  y},\quad
w\to e^{ \textup{i}  w}.
$$
Introduce also the exponential form of the parameters
$$
r= e^{-\alpha}, \quad s= e^{\alpha+\gamma-\eta}, \quad
t= e^{-\gamma},
$$
so that the balancing condition $r^2s^2t^2=1$ is satisfied automatically.
Finally, denote
\begin{equation}
W(\alpha;x,u):=\Delta (e^{\alpha-\eta}; e^{ \textup{i}  x},e^{\textup{i}  u}).
\label{weight}\end{equation}
Then relation (\ref{mstr}) can be rewritten as
\begin{eqnarray}\nonumber
&& \int_{0}^{2\pi} S(u;p,q)W(\eta-\alpha;x,u)W(\alpha+\gamma;y,u)W(\eta-\gamma;w,u)du
\\ && \makebox[2em]{}
=\chi(\alpha, \gamma;p,q)W(\alpha;y,w)W(\eta-\alpha- \gamma;x,w)
W(\gamma;x,y),
\label{astr}\end{eqnarray}
where
\begin{eqnarray}\label{S}
&& S(u;p,q)=\frac{(p;p)_\infty (q;q)_\infty}{4\pi}
\theta(e^{2 \textup{i}  u};p)\theta(e^{-2 \textup{i}  u};q),
\\  \makebox[0em]{}
&& \chi(\alpha,\gamma;p,q)= \Gamma(r^2,s^2,t^2;p,q).
\label{norms} \end{eqnarray}

As observed in \cite{BS}, equality (\ref{astr}) is nothing else than the star-triangle
relation playing an important role for solvable models of
statistical mechanics. It is symbolized by figure 1 given below, where
the black vertex of the star-shaped figure on the left-hand side means the
integration over $u$-variable with the weight $S(u)$, and $W$-weights are
associated with the edges connecting the black vertex with white ones.
On the right-hand side one has the product of three $W$-weights connecting
only white vertices.

\begin{figure}[h]
\unitlength=1mm
\makebox(70,25)[cc]{\psfig{file=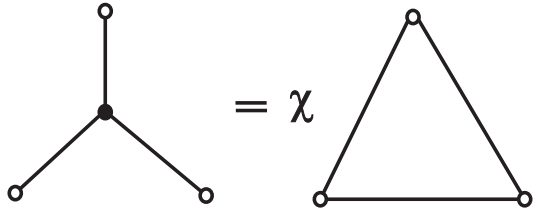,width=70mm}}
\caption{The star-triangle relation.}
\end{figure}

Suppose we have a two dimensional square lattice
with spin variables $a, b, c,\ldots$ sitting at vertices. One associates the
self-interaction energy  $S(a)$ with each spin (vertex). For
each horizontal bond connecting spins $a$ and $b$ the energy contribution
is given by the Boltzmann weight $W_{fg}(a,b)$, and the energy contribution
from each vertical bond connecting spins $b$ and $d$ is given by the weight
${\overline W}_{fg}(b,d)$. The variables $f$ and $g$ are called rapidities.
Then, as described in detail by Baxter in \cite{bax_ss,bax}, the general STR for
these quantities have the following functional equations form:
\begin{eqnarray}\nonumber && \makebox[-2em]{}
\sum_{d}S(d){\overline W}_{fg}(d,b) W_{fh}(c,d){\overline W}_{gh}(a,d)
=R_{fgh}W_{fg}(c,a){\overline W}_{fh}(a,b)W_{gh}(c,b),
\\ && \makebox[-2em]{}
\sum_{d}S(d){\overline W}_{fg}(b,d) W_{fh}(d,c){\overline W}_{gh}(d,a)
=R_{fgh}W_{fg}(a,c){\overline W}_{fh}(b,a)W_{gh}(b,c).
\label{str}\end{eqnarray}
The second equation is satisfied automatically if the Boltzmann weights are
symmetric in spin variables
$$
W_{fg}(a,b)=W_{fg}(b,a), \qquad {\overline W}_{fg}(a,b)={\overline W}_{fg}(b,a).
$$
Usually the normalization constants factorize, $R_{fgh}=r_{gh}r_{fg}/r_{fh}$.
Then the weights satisfy the unitarity relation of the form
$$
\sum_{d}S(d){\overline W}_{fg}(a,d) {\overline W}_{gf}(d,b)
=\frac{r_{fg}r_{gf}}{S(a)}\delta_{ab}
$$
and the reflection equation  $W_{fg}(a,b)W_{gf}(a,b)=1$.

A subclass of solutions of \eqref{str} emerges from the weights
depending only on differences of the rapidities,
\begin{equation}
W_{fg}(a,b)=W(f-g;a,b),\quad {\overline W}_{fg}(a,b)=W(\eta-f+g;a,b),
\label{W-ansatz}\end{equation}
where the parameter $\eta$ is called the crossing parameter.
Then the precise identification of equality \eqref{astr} with \eqref{str}
is reached after setting $\alpha=f-g,\ \gamma=g-h$
(so that $f-h=\alpha+\gamma$), equating $S(d)$ to $S(u;p,q)$ and
$R_{fgh}$ to $\chi(\alpha,\gamma;p,q)$ functions, and
fixing appropriately the range of summation (integration) over the variable $d=u$.
We call \eqref{weight}, \eqref{S}, \eqref{norms} the elliptic beta integral
STR solution. As shown in \cite{BS}, it generalizes many known solvable models
of statistical mechanics \cite{bax:book}: the Ising model, Ashkin-Teller, chiral Potts,
Fateev-Zamolodchikov $Z_N$-model, Kashiwara-Miwa and Faddeev-Volkov models.
Moreover, as will be shown below, it comprises also a new Faddeev-Volkov
type integrable system with continuous spins.

\begin{figure}[h]
\unitlength=1mm
\makebox(90,30)[cc]{\psfig{file=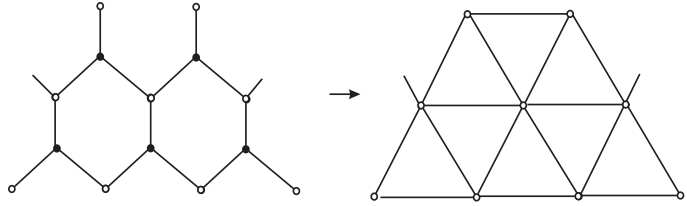,width=90mm}}
\caption{A honeycomb-triangluar lattice transformation induced by the
star-triangle relation.}
\end{figure}

There is direct relation between spin systems on lattices of
three types --- the honeycomb, triangular, and rectangular lattices.
Indeed, one can start from the honeycomb lattice, as depicted
on the left-hand side of figure 2.
Applying the star-triangle transformation to each black vertex
one transforms the whole honeycomb lattice to the triangular one
\cite{W}. In a similar way, one can apply STR to each white
vertex and obtain another triangular lattice having only black vertices.
This is quite evident and does not require further explanations.
However, further transformation of the triangular lattice to the
square one is more tricky.

Consider the left-hand side of figure 3.
Take the horizontal line in the middle of the drawn piece of the lattice.
Pick up the triangles above and below it which intersect only at one point
lying on this line (they are shown in bold lines).
Apply to them the triangle-star relation replacing them by stars
and continue this procedure up and down line-by-line of the resulting lattice.
As a result, one obtains eventually the square lattice. Taking into account the
nontrivial $\chi$-multiplier in STR, one can thus connect partition functions
of the square lattice model to the partition functions of two other types
of models.

\begin{figure}[h]
\unitlength=1mm
\makebox(90,45)[cc]{\psfig{file=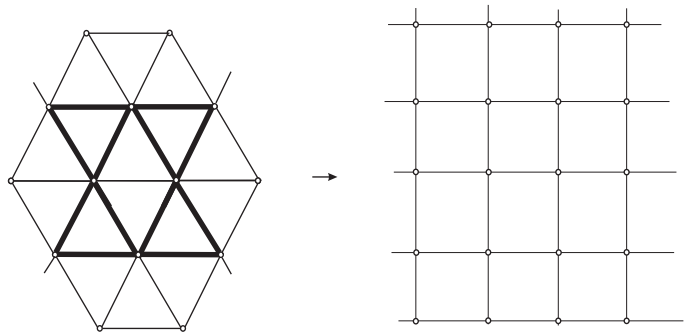,width=90mm}}
\caption{A triangluar-rectangular lattice transformation induced by the
star-triangle relation.}
\end{figure}

In \cite{BS}, the parameters $x$ and $u$ in \eqref{weight} were considered as
true spin variables. However, because of the $x\to -x$ and $u\to -u$ symmetries,
the Boltzmann weights $W$ and $S$ depend on their trigonometric combinations.
Therefore one can count as the true
spin variables  $U=\cos u,\, X=\cos x,$ etc, with their
values ranging from -1 to 1. The change of the variables in the measure
is elementary
$$
\int_{\mathbb T}f\Big(\frac{1}{2}(z+z^{-1})\Big)\frac{dz}{\textup{i} z}
=\int_0^{2\pi}f(\cos u)du=2\int_{-1}^1f(U)\frac{dU}{\sqrt{1-U^2}}.
$$

The Boltzmann weight $W(\alpha;x,u)$ satisfies the reflection symmetry
$$
W(\alpha;x,u)W(-\alpha;x,u)=1,
$$
 following from the reflection equation
for the elliptic gamma function. In terms of the spin variables
$X=\cos x$ and $Y=\cos y$ the unitarity relation takes the form
\begin{eqnarray}\nonumber && \makebox[0em]{}
\int_{-1}^1 S(u;p,q) W(\eta-\alpha;x,u)W(\eta+\alpha;y,u)
\frac{dU}{\sqrt{1-U^2}}
\\ && \makebox[4em]{}
=\frac{\Gamma(e^{2\alpha},e^{-2\alpha};p,q)}{S(x;p,q)}\sqrt{1-X^2}\, \delta(X-Y).
\label{unitarity}\end{eqnarray}
This equality has been established by the author in \cite{spi:con}.
Note that positivity of the Boltzmann weights $S(u;p,q)$ and $W(\alpha;x,u)$
corresponds to the unitarity of the elliptic modular double representations
\cite{spi:con}. In particular, they are positive for $x,u\in[0,2\pi]$,
real $\alpha$ such that $|\sqrt{pq}e^{\alpha}|<1$, and
$$
1)\ \ p^*=p,\quad q^*=q,  \qquad\text{or}\qquad 2)\  \ p^*=q.
$$
At the level of superconformal indices, relations similar to
\eqref{unitarity} describe the Seiberg dualities for gauge field
theories with equal number of colors and flavors and the chiral symmetry breaking
\cite{SVprep}.

Relation (\ref{astr}) is not changed if one replaces $W$ and $\chi$ by
\begin{eqnarray}  \nonumber &&
\widetilde W(\alpha;x,u)=\frac{W(\alpha;x,u)}{m(\alpha)},\quad
\\ &&
\tilde \chi(\alpha, \gamma;p,q)=\frac{m(\alpha)m(\gamma)m(\eta-\alpha-\gamma)}
{m(\eta-\alpha)m(\eta-\gamma)m(\alpha+\gamma)}
\chi(\alpha, \gamma;p,q)
\label{renorm}\end{eqnarray}
for arbitrary normalizing factor $m(\alpha)$.

The star-triangle relation is one of the three known forms of the
Yang-Baxter equation. The second, probably the most popular form,
is the vertex type relation symbolically written in terms of the
$R$-matrices as
\begin{equation}\label{YB}
{\bf R}^{(12)}(\lambda) {\bf R}^{(13)}(\lambda+\mu) {\bf R}^{(23)}(\mu)
={\bf R}^{(23)}(\mu) {\bf R}^{(13)}(\lambda+\mu){\bf R}^{(12)}(\lambda),
\end{equation}
where $\lambda$ and $\mu$ are spectral parameters. The third type
is referred to as the IRF (interaction around the face) Yang-Baxter equation.
The star-star relation, which was discussed in detail in \cite{bax_ss},
belongs to the latter type of equations and has the form
\begin{eqnarray}\nonumber
&& \sum_g S(g) W_1(a,g)W_2(b,g)W_3(c,g)W_4(d,g)
\\ && \makebox[2em]{}
=R\frac{m(b,c)p(a,b)}{m(a,d)p(c,d)}
\sum_g S(g) W_1'(a,g)W_2'(b,g)W_3'(c,g)W_4'(d,g),
\label{ssr}\end{eqnarray}
where $W_j(a,b), W'_j(a,b), m(b,c), p(a,b)$ are two-spin Boltzmann weights
and $S(g)$ is the spin
self-interaction weight (it was omitted in formula (1.1) of \cite{bax_ss}).
The left-hand side
can be interpreted as an elementary partition
function for a system of four spins $a, b, c, d$
sitting in four square vertices connected by
edges to the spin $g$ sitting in the square
center, and the summation is going over the values of the central spin,
see figure 4 below. The right hand side
has a similar interpretation of a statistical sum multiplied by the additional Boltzmann
weights associated with opposite edges of the square $(a,b,c,d)$.
Formula \eqref{ssr} can be thought of as a generalized Kramers-Wannier duality
transformation \cite{KW,W}.

\begin{figure}[h]
\unitlength=1mm
\makebox(90,30)[cc]{\psfig{file=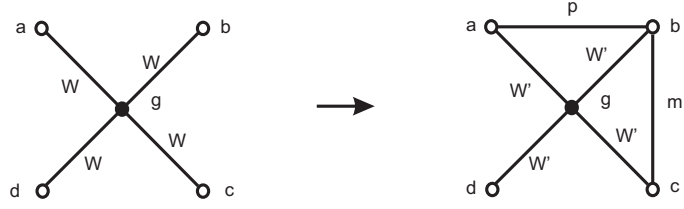,width=90mm}}
\caption{A star-star relation for the square lattice.
Additional Boltzmann weights $p$ and $m$ are indicated by edges
connecting corresponding vertices on the right-hand side.}
\end{figure}

Relation \eqref{ssr} should be compared with the $V$-function
symmetry transformations written in the form \eqref{t1}, \eqref{t2}, and \eqref{t3}.
Some of them coincide with \eqref{ssr} after appropriate identifications
of the Boltzmann weights. For instance, equation \eqref{t1} corresponds to the choice
\begin{eqnarray*}
&& W_1(a,g)=\Delta(\alpha;x,g), \quad\ \ \ \ \ W_1'(a,g)= \Delta(\sqrt{pq}\beta^{-1};x,g),
\\
&& W_2(b,g)=\Delta(\beta;y,g), \quad \ \ \ \ \ \, W_2'(b,g)= \Delta(\sqrt{pq}\alpha^{-1};y,g),
\\
&& W_3(c,g)=\Delta(\sqrt{pq}\gamma;w,g), \quad  W_3'(c,g)= \Delta(\delta^{-1};w,g),
\\
&& W_4(d,g)=\Delta(\sqrt{pq}\delta;z,g), \quad\,  W_4'(d,g)= \Delta(\gamma^{-1};z,g),
\end{eqnarray*}
where $\alpha\beta\gamma\delta=1$, $g$ is the integration variable
for the $V$-function, and $S(g)=\kappa/\Gamma(g^{\pm2};p,q)$.
Other factors have the form
\begin{eqnarray}\nonumber
&&
R=\frac{\Gamma(\alpha^2,\beta^2;p,q)}
{\Gamma(\gamma^{-2},\delta^{-2};p,q)}, \qquad m(b,c)=m(a,d)=1,
\\ &&
p(a,b)=\Delta(\alpha\beta;x,y), \qquad \ p(c,d)=\Delta(\alpha\beta;w,z).
\label{ssr1}\end{eqnarray}

A similar interpretation is valid for relation
\eqref{t2}. It corresponds to the choice
\begin{eqnarray*}
&& W_1(a,g)=\Delta(\alpha;x,g), \quad\ \ \ \ \ \ \ \ \ \ W_1'(a,g)= \Delta(\beta;x,g),
\\
&& W_2(b,g)=\Delta(\sqrt{pq}\gamma;w,g), \quad \  \ \ \ \,
   W_2'(b,g)= \Delta(\sqrt{pq}\delta;w,g),
\\
&& W_3(c,g)=\Delta(\beta;y,g), \quad  \ \ \ \ \ \ \ \ \ \ W_3'(c,g)= \Delta(\alpha;y,g),
\\
&& W_4(d,g)=\Delta(\sqrt{pq}\delta;z,g), \quad\ \ \ \ \
W_4'(d,g)= \Delta(\sqrt{pq}\gamma;z,g),
\end{eqnarray*}
where, again, $\alpha\beta\gamma\delta=1$ and $S(g)=\kappa/\Gamma(g^{\pm2};p,q)$.
As to other factors, $R=1$ and
\begin{eqnarray}\nonumber
&&
 m(b,c)=\Delta(\sqrt{pq}\beta\gamma;y,w),\qquad \
m(a,d)=\Delta(\sqrt{pq}\beta\gamma;x,z),
\\ &&
p(a,b)=\Delta(\sqrt{pq}\alpha\gamma;x,w), \qquad \ \,
p(c,d)=\Delta(\sqrt{pq}\alpha\gamma;y,z).
\label{ssr2}\end{eqnarray}

There are three star-star relations for the Ising type models
listed in \cite{bax_ss} as equations (2.16), (5.1), and (5.2).
Our first option  \eqref{ssr1} corresponds to relation
(5.2) in \cite{bax_ss}. Relations (2.16) and (5.2) in \cite{bax_ss}
are obtained from each other by a
reflection with respect to the lattice square diagonal $(b,d)$.
Our second option \eqref{ssr2} corresponds to relation (5.1) in \cite{bax_ss}
with nonconstant $p$- and $m$-weights.
However, we have the third nontrivial form of the symmetry
transformation for the $V$-function \eqref{t3}. It corresponds to
a more complicated type of the star-star relation
\begin{eqnarray}\label{ssr_gen}
&& \sum_g S(g) W_1(a,g)W_2(b,g)W_3(c,g)W_4(d,g)
\\ && \makebox[2em]{}
=R\frac{m(b,c)p(a,b)t(a,c)}{m(a,d)p(c,d)t(b,d)}
\sum_g S(g) W_1'(a,g)W_2'(b,g)W_3'(c,g)W_4'(d,g),
\nonumber\end{eqnarray}
where $t(a,c)$ is a new diagonal Boltzmann weight. Explicitly, we have
\begin{eqnarray*}
&& W_1(a,g)=\Delta(\alpha;x,g), \quad\ \ \ \ \ W_1'(a,g)= \Delta(\sqrt{pq}\alpha^{-1};x,g),
\\
&& W_2(b,g)=\Delta(\beta;y,g), \quad \ \ \ \ \ \, W_2'(b,g)= \Delta(\sqrt{pq}\beta^{-1};y,g),
\\
&& W_3(c,g)=\Delta(\sqrt{pq}\gamma;w,g), \quad  W_3'(c,g)= \Delta(\gamma^{-1};w,g),
\\
&& W_4(d,g)=\Delta(\sqrt{pq}\delta;z,g), \quad\,  W_4'(d,g)= \Delta(\delta^{-1};z,g),
\end{eqnarray*}
where $\alpha\beta\gamma\delta=1$. Other factors in \eqref{ssr_gen} are
$R=\Gamma(\alpha^2,\beta^2;p,q)/\Gamma(\gamma^{-2},\delta^{-2};p,q)$
and
\begin{eqnarray*}
&&
 m(b,c)=\Delta(\sqrt{pq}\beta\gamma;y,w),\qquad
m(a,d)=\Delta(\sqrt{pq}\beta\gamma;x,z),
\\ &&
p(a,b)=\Delta(\alpha\beta;x,y), \qquad \ \ \ \ \ \
p(c,d)=\Delta(\alpha\beta;w,z),
\\ &&
t(a,c)=\Delta(\sqrt{pq}\alpha\gamma;x,w), \qquad \
t(b,d)=\Delta(\sqrt{pq}\alpha\gamma;y,z).
\end{eqnarray*}
Perhaps, this type of the star-star relation was not considered in the literature
before. Note that all such relations represent symmetry groups of
the partition functions. In the case of $V$-function this is $W(E_7)$,
i.e. one has much bigger symmetry than that seen explicitly
in the chosen spin system interpretation.
We have described thus a new (elliptic hypergeometric) class of solutions of
the star-star relation which should lead to
new solvable models of statistical mechanics similar to the checkerboard Ising model.
Known systems of such type were investigated in detail in \cite{BaSt}. A natural
general conclusion from our consideration is that the symmetry of STR can be
richer than a direct sum of symmetries of the Boltzmann weights and
the lattice.

\section{A hyperbolic beta integral STR solution}

We describe now another solution of the star-triangle relation
associated with the modified form of the elliptic beta integral
when one of the bases $p$ or $q$ can lie on the
unit circle \cite{ds:unit}. It simplifies also consideration of
the degeneration limits to $q$-beta integrals of the Mellin-Barnes type
(hyperbolic beta integrals).

First we describe the modified elliptic gamma function introduced in
\cite{spi:theta}. It is convenient to use additive notation
and introduce three pairwise incommensurate
quasiperiods $\omega_1$, $\omega_2$, $\omega_3$ together with the
definitions
\begin{eqnarray}\nonumber
&& q=e^{2\pi \textup{i} \frac{\omega_1}{\omega_2}}, \qquad \ \
p=e^{2\pi \textup{i}\frac{\omega_3}{\omega_2}},\qquad \ \
r=e^{2\pi \textup{i}\frac{\omega_3}{\omega_1}},
\\
&& \tilde q =e^{-2\pi \textup{i} \frac{\omega_2}{\omega_1}}, \qquad \tilde
p=e^{-2\pi \textup{i} \frac{\omega_2}{\omega_3}},\qquad \tilde r=e^{-2\pi
\textup{i}\frac{\omega_1}{\omega_3}}. \label{ell-bases}\end{eqnarray}
Here $\tilde q, \tilde p,$ and $\tilde r$ are particular ($\tau\to -1/\tau$)
modular transformations of $q,p,$ and $r$. Assume that
$\text{Im}(\omega_1/\omega_2),\text{Im}(\omega_3/\omega_1),
\text{Im}(\omega_3/\omega_2)>0$, or $|q|, |p|, |r|<1$.
Then the modified elliptic gamma function is constructed as a product
of two elliptic gamma functions
\begin{eqnarray} \nonumber &&
G(u;\omega_1,\omega_2,\omega_3)=\Gamma(e^{2\pi \textup{i} \frac{u}{\omega_2}};p,q)
\Gamma(re^{-2\pi \textup{i} \frac{u}{\omega_1}};r,\tilde q)
\\ && \makebox[6,7em]{}
=e^{-\frac{\pi \textup{i} }{3} B_{3,3}(u;\mathbf{\omega})}\Gamma(e^{-2\pi
\textup{i} \frac{u}{\omega_3}}; \tilde r, \tilde p),
\label{ell-d}\end{eqnarray}
where $B_{3,3}(u;\mathbf{\omega})$ is the third diagonal Bernoulli polynomial
(for the general definition of such polynomials, see Appendix A),
$$
B_{3,3}\left(u+\sum_{n=1}^3\frac{\omega_n}{2};\mathbf{\omega}\right)
=\frac{u(u^2-\frac{1}{4}\sum_{k=1}^3\omega_k^2)}{\omega_1\omega_2\omega_3}.
$$
The $G(u;\mathbf{\omega})$-function satisfies the following system of three linear
difference equations of the first order
\begin{eqnarray*}
&& G(u+\omega_1;\mathbf{\omega})=\theta(e^{2\pi
\textup{i}\frac{u}{\omega_2}};p) G(u;\mathbf{\omega}),
\\
&& G(u+\omega_2;\mathbf{\omega})=\theta(e^{2\pi
\textup{i}\frac{u}{\omega_1}};r) G(u;\mathbf{\omega}),
 \\
&& G(u+\omega_3;\mathbf{\omega})= e^{-\pi \textup{i} B_{2,2}(u;\mathbf{\omega})}
 G(u;\mathbf{\omega}),
\end{eqnarray*}
where $B_{2,2}(u;\mathbf{\omega})$ is the second diagonal Bernoulli polynomial,
$$
B_{2,2}(u;\mathbf{\omega})=\frac{u^2}{\omega_1\omega_2}
-\frac{u}{\omega_1}-\frac{u}{\omega_2}+
\frac{\omega_1}{6\omega_2}+\frac{\omega_2}{6\omega_1}+\frac{1}{2}.
$$
The second equality in (\ref{ell-d}) follows from the fact that both
expressions for $G(u;\mathbf{\omega})$ satisfy the above set of
equations and the normalization
$G(\frac{1}{2}\sum_{k=1}^3\omega_k;\mathbf{\omega})=1$.

It is easy to see that $G(u;{\bf \omega})$ is well defined for
$|p|,|r|<1$ and $|q|\le1$, the $|q|=1$ case being permitted in difference from
the $\Gamma(z;p,q)$-function. Evidently, we have the symmetry relation
$$
G(u;\omega_1,\omega_2,\omega_3)=G(u;\omega_2,\omega_1,\omega_3)
$$
and the reflection equation
$$
G(a;{\bf \omega})G(b;{\bf \omega})=1,\quad a+b=\sum_{k=1}^3\omega_k.
$$

For $\text{Im}(\omega_{1}/\omega_2)>0$, we can take the limit
$\omega_3\to \infty$ in such a way that
$$
\text{Im}(\omega_{3}/\omega_1), \quad
\text{Im}(\omega_{3}/\omega_2)\to +\infty
$$
and $p,r\to 0$. Then,
\begin{equation}
\lim_{p,r\to 0} G(u;\mathbf{\omega})
=\gamma(u;\omega_1,\omega_2)
= \frac{(e^{2\pi \textup{i} u/\omega_1}\tilde q; \tilde q)_\infty}
{(e^{2\pi \textup{i}  u/\omega_2}; q)_\infty}.
\label{2d-sin}\end{equation}
For $\text{Re}(\omega_1), \text{Re}(\omega_2)>0$ and
$0<\text{Re}(u)<\text{Re}(\omega_1+\omega_2)$ this $\gamma$-function
has the following integral representation
\be
\gamma(u;\omega_1,\omega_2)=
\exp\left(-\int_{\R+\textup{i} 0}\frac{e^{ux}}
{(1-e^{\omega_1 x})(1-e^{\omega_2 x})}\frac{dx}{x}\right),
\label{mod-q-gamma}\ee
which shows that $\gamma(u;\omega_1,\omega_2)$ is a meromorphic function of $u$
even for $\omega_1/\omega_2>0$, when $|q|=1$ and the infinite product representation
\eqref{2d-sin} is not valid any more.
The inversion relation for this function has the form
$$
\gamma(u;\omega_1,\omega_2)\gamma(\omega_1+\omega_2-u;\omega_1,\omega_2)
=e^{\pi\textup{i}B_{2,2}(u;\mathbf{\omega})}.
$$
For more details on this function see Appendix A.

Let $\text{Im} (\omega_1/\omega_2)\geq 0$ and $\text{Im}
(\omega_3/\omega_1), \text{Im} (\omega_3/\omega_2)>0$, and
let six complex parameters $g_k$, $k=1,\ldots,6$, satisfy
the constraints $\text{Im}(g_k/\omega_3)<0$ and
\begin{equation}
\sum_{k=1}^6 g_k=\omega_1+\omega_2+\omega_3.
\label{bal}\end{equation}
Then  \cite{ds:unit},
\begin{equation}
\int_{-\omega_3/2}^{\omega_3/2} \frac{\prod_{k=1}^6 G(g_k\pm
u;\mathbf{\omega})} {G(\pm 2u;\mathbf{\omega})} du
 = \tilde\kappa\, \prod_{1\leq k<l\leq 6}G(g_k+g_l;\mathbf{\omega}),
\label{circle-int}\end{equation}
where the integration goes along the straight line segment
connecting points $-\omega_3/2$ and $\omega_3/2$, and
\begin{equation}
\tilde\kappa= \frac{-2\omega_2(\tilde q;\tilde q)_\infty}
{(q;q)_\infty(p;p)_\infty(r;r)_\infty}. \label{kappa}
\end{equation}
Here and below we use the shorthand notation
$$
G(a\pm b;\mathbf{\omega}):=G(a+b,a-b;\mathbf{\omega})
:=G(a+b;\mathbf{\omega})G(a-b;\mathbf{\omega}).
$$

The proof of equality \eqref{circle-int} is rather simple. It is necessary to
substitute in it the second form of $G(u;\mathbf{\omega})$-function
\eqref{ell-d}, check that all exponential factors cancel and, after a change
of notation, the formula reduces to the standard elliptic beta integral.

Let us introduce the crossing parameter
$$
\eta=-\frac{1}{2}\sum_{k=1}^3\omega_k
$$
and denote
\be
g_{1,2}=-\alpha\pm x,\quad g_{3,4}=\alpha+\gamma-\eta \pm y, \quad
g_{5,6}=-\gamma\pm w,
\label{h-pars1}\ee
so that the balancing condition \eqref{bal} is  satisfied automatically.
Introduce also the modified Boltzmann weight, or the kernel for
the modified form of the integral transformation (\ref{trn}),
$$
W'(\alpha;x,u)=G(\alpha-\eta\pm x\pm u;\mathbf{\omega}).
$$
Then relation (\ref{circle-int}) can be rewritten as
\begin{eqnarray}\nonumber
&& \int_{-\omega_3/2}^{\omega_3/2} \phi(u;\mathbf{\omega})
W'(\eta-\alpha;x,u)W'(\alpha+\gamma;y,u)
W'(\eta-\gamma;w,u)du
\\ && \makebox[2em]{}
=\chi(\alpha, \gamma;\mathbf{\omega})W'(\alpha;y,w)
W'(\eta-\alpha- \gamma;x,w)W'(\gamma;x,y),
\label{mastr}\end{eqnarray}
where
\begin{eqnarray}\nonumber
&& \phi(u;\mathbf{\omega})=\frac{1}{\tilde\kappa G(\pm 2u;\mathbf{\omega})}
= \frac{1}{\tilde\kappa}
e^{-\pi \textup{i}  B_{2,2}(2u;\omega_1,\omega_2)}
\theta(e^{-4\pi \textup{i}  u/\omega_2};p)\theta(e^{-4\pi \textup{i}  u/\omega_1};r),
\\  \makebox[0em]{}
&& \chi(\alpha,\gamma;\mathbf{\omega})=G(-2\alpha,-2\gamma,
2\alpha+2\gamma-2\eta;\mathbf{\omega}).
\label{normadd2} \end{eqnarray}

Substituting the second form of the modified elliptic gamma function,
we find
\begin{eqnarray}\nonumber &&
W'(\alpha;x,u)=\exp\left(-\frac{4\pi \textup{i} }{3}\Big(B_{3,3}(\alpha-\eta;\mathbf{\omega})
+\frac{3\alpha(x^2+u^2) }{\omega_1\omega_2\omega_3}\Big)\right)
\\ && \makebox[6em]{} \times
\Delta\left(e^{2\pi \textup{i} (\eta-\alpha)};\frac{x}{\omega_3},\frac{u}{\omega_3};
\tilde p,\tilde r\right).
\label{modweight}\end{eqnarray}
We see that this Boltzmann weight is obtained from \eqref{weight} after a reparametrization
of variables and multiplication by an exponential of
a quadratic polynomial of the spin variables.
This means that there exists a nontrivial symmetry transformation of
the star-triangle relation modifying its solutions in the described way.

The distinguished property of the modified elliptic
beta integral is that it is well defined for $|q|=1$.
Therefore the limit $\omega_3\to\infty$ leads to $q$-beta integrals
well defined in this regime as well.
Let $\text{Re}(\omega_1),\ \text{Re}(\omega_2)>0$. Then, for
$\omega_3\to +\textup{i}\infty$, one has $p,\, r\to 0$ and $G(u;\mathbf{\omega})$
goes to $\gamma(u;\omega_1, \omega_2)$-function. Let us substitute
$g_6=\sum_{k=1}^3\omega_k -A$ in formula \eqref{circle-int}, where $A=\sum_{k=1}^5g_k$,
and apply the inversion formula to the corresponding modified elliptic gamma function.
Then the formal limit $\omega_3\to +\textup{i}\infty$ reduces this
integration formula to
\begin{eqnarray}
&& \int_{-\textup{i}\infty}^{+\textup{i}\infty}
\frac{\prod_{j=1}^5\gamma(g_k\pm u;\mathbf{\omega}) }
{\gamma(\pm 2u,A\pm u;\mathbf{\omega}) } du
=-2\omega_2\frac{(\tilde q;\tilde q)_\infty}{(q;q)_\infty}
\frac{\prod_{1\leq j<k\leq 5}\gamma(g_j+g_k;\mathbf{\omega})}
{\prod_{k=1}^5\gamma(A-g_k;\mathbf{\omega})},
\label{stok1}\end{eqnarray}
where the integration contour is the straight line for $\text{Re}(g_k)>0$
or the Mellin-Barnes type contour, if these restrictions for parameters
are violated. Let us remind also that
$$
\frac{(\tilde q;\tilde q)_\infty}{(q;q)_\infty}
=\sqrt{-\textup{i}\frac{\omega_1}{\omega_2}}
e^{\frac{\pi \textup{i}}{12}\left(\frac{\omega_1}{\omega_2}
+\frac{\omega_2}{\omega_1}\right)},
$$
where $\sqrt{-\textup{i}}=e^{-\pi \textup{i}/4}$ since for
$\omega_1/\omega_2=\textup{i} a,\, a>0$,
the square root should be positive.

Let us introduce parameter $g_6$ anew (it should not be confused with
the previous variable $g_6$ which we have eliminated) using the condition
\begin{equation}
\sum_{k=1}^6g_k=\omega_1+\omega_2
\label{bal_hyper}\end{equation}
(note the difference with \eqref{bal}).
Now we can apply the inversion formula to $\gamma$-functions
to move some of them from the denominator of the integral kernel
to its numerator.
It is convenient here to define the hyperbolic gamma function
$\gamma^{(2)}(u)$:
\begin{equation}
\gamma^{(2)}(u;\mathbf{\omega})= e^{-\frac{\pi\textup{i}}{2}
B_{2,2}(u;\mathbf{\omega}) } \gamma(u;\mathbf{\omega}).
\label{barg}\end{equation}
Then, after the change of the integration variable $u=\textup{i}z$,
the integral \eqref{stok1} takes the compact form
\begin{eqnarray}
&& \int_{-\infty}^\infty
\frac{\prod_{j=1}^6\gamma^{(2)}(g_k\pm \textup{i}z;\mathbf{\omega}) }
{\gamma^{(2)}(\pm 2 \textup{i}z;\mathbf{\omega}) } dz
=2\sqrt{\omega_1\omega_2}
\prod_{1\leq j<k\leq 6}\gamma^{(2)}(g_j+g_k;\mathbf{\omega}).
\label{stok2}\end{eqnarray}

Validity of the described limit $\omega_3\to\infty$ at the level of
integrals was rigorously justified in
\cite{rai:limits} using a slightly different notation.
Integral \eqref{stok2} was proven first (using a different approach)
by Stokman \cite{stok} who called it the hyperbolic beta integral.
 We followed the formal limiting procedure suggested in \cite{ds:unit}.

Similar to  \eqref{h-pars1}, let us fix the parameters as
\be
g_{1,2}=-\alpha\pm \textup{i}x,\quad g_{3,4}=\alpha+\gamma-\eta \pm \textup{i}y, \quad
g_{5,6}=-\gamma\pm \textup{i} w
\label{h-pars2}\ee
with the crossing parameter $\eta=-(\omega_1+\omega_2)/2$.
Then formula \eqref{stok2} can be rewritten as a star-triangle relation
\begin{eqnarray}\nonumber
&& \int_{-\infty}^\infty S(z)W(\eta-\alpha;x,z)W(\alpha+\gamma;y,z)W(\eta-\gamma;w,z)dz
\\ && \makebox[2em]{}
=\chi(\alpha, \gamma)W(\alpha;y,w)W(\eta-\alpha- \gamma;x,w)
W(\gamma;x,y),
\label{Hstr}\end{eqnarray}
where
\begin{eqnarray}\nonumber
&& W(\alpha;x,z)=\gamma^{(2)}(\alpha-\eta\pm \textup{i}x\pm \textup{i}z;\mathbf{\omega}),
\\  \makebox[0em]{}
&& S(z)=\frac{1}{2\sqrt{\omega_1\omega_2}\gamma^{(2)}(\pm 2\textup{i} z;\mathbf{\omega})}
=\frac{2\sinh \frac{2\pi z}{\omega_1}\sinh \frac{2\pi z}{\omega_2}}
{\sqrt{\omega_1\omega_2}},
\nonumber\\  \makebox[0em]{}
&& \chi(\alpha,\gamma)=\gamma^{(2)}(-2\alpha,
-2\gamma,2\alpha+2\gamma-2\eta;\mathbf{\omega}).
\label{modSTR}\end{eqnarray}
These Boltzmann weights are positive for real $x, z, \alpha$,
$\eta<\alpha<-\eta$, $\eta<0$, and
either real $\omega_{1,2}$ or $\omega_1^*=\omega_2$.

Consider a particular reduction of integration formula \eqref{stok2}.
For this we replace parameters $g_j\to g_j+\textup{i} \mu $, $j=1,2,3,$ and
$g_j\to g_j-\textup{i} \mu $, $j=4,5,6.$  Since the integrand is symmetric in $z$
we can rewrite the left-hand side as
\begin{eqnarray*} &&
2\int_0^\infty\frac{\prod_{j=1}^3\gamma^{(2)}(g_j+\textup{i}\mu\pm \textup{i}z,
g_{j+3}-\textup{i}\mu\pm \textup{i}z;\mathbf{\omega})}
{\gamma^{(2)}(\pm 2\textup{i}z;\mathbf{\omega})} dz
\\ && \makebox[2em]{}
=2\int_{-\mu}^\infty\prod_{j=1}^3\gamma^{(2)}(g_{j}-\textup{i}z,
g_{j+3}+\textup{i}z;\mathbf{\omega})\rho_1(z)\rho_2(z)dz,
\end{eqnarray*}
where
$$
\rho_1(z)=\frac{e^{-2\pi(z+\mu)(\omega_1^{-1}+\omega_2^{-1})}}
{\gamma^{(2)}(\pm 2\textup{i}(z+\mu);\mathbf{\omega})}
\stackreb{\to}{\mu\to+\infty} 1
$$
and
\begin{eqnarray*} &&
\rho_2(z)=e^{2\pi(z+\mu)(\omega_1^{-1}+\omega_2^{-1})}
\prod_{j=1}^3\gamma^{(2)}(g_j+2\textup{i}\mu+ \textup{i}z,
g_{j+3}-2\textup{i}\mu-\textup{i}z;\mathbf{\omega})
\\ && \makebox[-2em]{}
\stackreb{\to}{\mu\to+\infty} e^{\frac{\pi}{\omega_1\omega_2}\Big(-2\mu(\omega_1+\omega_2)
+\frac{\textup{i}}{2}\sum_{j=1}^3\Big(g_{j+3}^2-g_j^2
+(g_j-g_{j+3})(\omega_1+\omega_2)\Big)\Big)}(1+o(1)).
\end{eqnarray*}
On the right-hand side we find
$$
2\sqrt{\omega_1\omega_2}\prod_{j=1}^3\prod_{k=4}^6
\gamma^{(2)}(g_j+g_k;\mathbf{\omega})\rho_3(g),
$$
where
\begin{eqnarray*} &&
\rho_3(g)=\prod_{1\leq j<k\leq 3}
\gamma^{(2)}(g_j+g_k+2\textup{i}\mu,g_{j+3}+g_{k+3}-2\textup{i}\mu;\mathbf{\omega})
\\ && \makebox[-2em]{}
\stackreb{\to}{\mu\to+\infty}
e^{\frac{\pi\textup{i}}{2} \sum_{1\leq j<k\leq 3}\Big(
B_{2,2}(g_{j+3}+g_{k+3}-2\textup{i}\mu)
-B_{2,2}(g_{j}+g_{k}+2\textup{i}\mu)\Big)}(1+o(1)).
\end{eqnarray*}
One can check that the leading asymptotics of $\rho_3(g)$ coincides with that
of the $\rho_2(z)$-function.
Taking the limit $\mu\to+\infty$, which is uniform, one comes
to the following exact integration formula \cite{bult}
\begin{equation}
\int_{-\infty}^\infty\prod_{j=1}^3\gamma^{(2)}(g_{j}-\textup{i}z,
g_{j+3}+\textup{i}z;\mathbf{\omega})dz
=\sqrt{\omega_1\omega_2}\prod_{j=1}^3\prod_{k=4}^6
\gamma^{(2)}(g_j+g_k;\mathbf{\omega}),
\label{bult}\end{equation}
where $\sum_{k=1}^6g_k=\omega_1+\omega_2$.

Let us change notation for the integral parameters
\begin{eqnarray*}
&& g_1=-\alpha+\textup{i}x, \qquad g_2=\alpha+\gamma-\eta+\textup{i}y, \qquad
g_3=-\gamma+\textup{i}w,
\\ && \makebox[0em]{}
g_4=-\alpha-\textup{i}x, \qquad g_5=\alpha+\gamma-\eta-\textup{i}y, \qquad
g_6=-\gamma-\textup{i}w,
\end{eqnarray*}
where $\eta=-(\omega_1+\omega_2)/2$ is a crossing parameter; the
balancing condition $\sum_{k=1}^6g_k=-2\eta$ is satisfied then automatically.
Now we can rewrite equality \eqref{bult} as the star-triangle relation
\begin{eqnarray}\nonumber
&& \int_{-\infty}^\infty W(\eta-\alpha;x,z)W(\alpha+\gamma;y,z)W(\eta-\gamma;w,z)
\frac{dz}{\sqrt{\omega_1\omega_2}}
\\ && \makebox[3em]{}
=\chi\, W(\alpha;y,w)W(\eta-\alpha- \gamma;x,w)
W(\gamma;x,y),
\label{astrFV}\end{eqnarray}
where the Boltzmann weight is defined as
$$
W(\alpha;x,z)=\gamma^{(2)}(\alpha-\eta\pm \textup{i} (x-z);\mathbf{\omega})
$$
and the normalization constant is
\begin{equation}
\chi=\gamma^{(2)}(-2\alpha,-2\gamma,
2\alpha+2\gamma-2\eta;\mathbf{\omega}).
\label{normHBISTR}\end{equation}
Note that $W(\alpha;x,y)=W(\alpha;y,x)$ and $W(\alpha;x,y)>0$
in the same domain of parameters as before.
 Denoting $\omega=b$, $\omega_2=b^{-1}$,
and $\alpha=-(b+b^{-1})\theta/(2\pi)$ one can see that $W(\alpha;x,z)$ coincides with
the Boltzmann weight of the Faddeev-Volkov model \cite{FV,VF} denoted as
$ W_\theta(x-z)$ in \cite{BMS} (our $\eta$ differs by sign from the definition
chosen in \cite{BMS}) up to some normalization factor $F_\theta$.

We thus see that the Faddeev-Volkov model solution of the
star-triangle relation \cite{VF} is a particular case of our
hyperbolic beta integral STR solution \eqref{modSTR}.
The fact that the left-hand side of STR for the Faddeev-Volkov model
represents a particular limiting case of the elliptic beta integral
was known to the author already in 2008.
After seeing \cite{BMS} and understanding this fact, the author
was interested whether a similar interpretation exists
for the elliptic beta integral itself. However, this idea was not developed
further, partially because the origin of the normalizing factor $F_\theta$
given in  \cite{BMS} was not understood at that time. Fortunately,
Bazhanov and Sergeev have independently answered this question in \cite{BS}.

\section{Partition functions}

The partition function of a homogeneous two dimensional discrete spin system
on the square lattice with the Boltzmann weights $W(\alpha;u_i,u_j)$ \eqref{weight}
and $S(u_j)$ \eqref{S} has the form
$$
Z=\int \prod_{(ij)}W(\alpha;u_i,u_j)\prod_{(kl)}W(\eta-\alpha;u_k,u_l)
\prod_{m}S(u_m)du_m,
$$
where the first product is taken over the horizontal edges $(ij)$,
the second product goes over all vertical edges $(k,l)$,
and the third product (in $m$) is taken over all internal vertices of the
lattice.
Let us take the elliptic beta integral STR solution of \cite{BS}
and consider the contribution to $Z$ coming from a particular
vertex $u$ surrounded by the vertices $u_{1}, u_2, u_3, u_4$:
$$
\int_0^{2\pi}S(u)W(\alpha;u_1,u)W(\alpha;u,u_3)W(\eta-\alpha;u_2,u)W(\eta-\alpha;u,u_4)du.
$$
Substituting explicit expressions for the weights, one can easily see
that this integral is equal to the elliptic hypergeometric function
$V(t_1,\ldots,t_8;p,q)$ described above \eqref{eghf}
with the following restricted set of parameters
$$
\{t_1,t_2,t_3,t_4\}=\{e^{\alpha-\eta}e^{2\pi\textup{i}u_{1}},
e^{\alpha-\eta}e^{-2\pi\textup{i}u_{1}},
e^{\alpha-\eta}e^{2\pi\textup{i}u_{3}},
e^{\alpha-\eta}e^{-2\pi\textup{i}u_{3}}\},
$$
$$
\{t_5,t_6,t_7,t_8\}=\{e^{-\alpha}e^{2\pi\textup{i}u_{2}},
e^{-\alpha}e^{-2\pi\textup{i}u_{2}},
e^{-\alpha}e^{2\pi\textup{i}u_{4}},
e^{-\alpha}e^{- 2\pi\textup{i}u_{4}}\}.
$$
In total, there are 5 independent parameters, instead of 7 for generic $V$-function
(in addition to the bases $p$ and $q$).
Therefore we conclude that the full partition function $Z$ is given
by an elliptic hypergeometric integral constructed as a tower
of intertwined (restricted) elliptic analogues of the Gauss hypergeometric
function similar to the Bailey tree for integrals \cite{spi:bai}.

According to the general reflection method used in \cite{BS}, the leading
asymptotics of the partition function for two-dimensional $N\times M$ lattice
when its size goes to infinity, $N,M\to\infty$, has the form
$$
Z\stackreb{=}{N,M\to\infty} m(\alpha) ^{NM},
$$
where $m(\alpha)$ is the normalizing factor for Boltzmann weights
which guarantees that on the right-hand side of STR the $\tilde\chi$-multiplier
\eqref{renorm} is equal to unity, $\tilde \chi=1$.
This condition is satisfied if
\be
\frac{m(\alpha)}{m(\eta-\alpha)}\Gamma(e^{-2\alpha};p,q)=1,
\quad \text{or}\quad
m(\alpha+\eta)=\Gamma(e^{2\alpha};p,q)m(-\alpha).
\label{normeq}\ee

Let us introduce the function
\begin{equation}
M(x;p,q,t)=\exp\Big(\sum_{n\in\Z/\{0\}} \frac{(\sqrt{pqt}x)^n}{n(1-p^n)(1-q^n)(1+t^n)}\Big)
=\frac{\Gamma(xt\sqrt{pqt};p,q,t^2)}{\Gamma(x\sqrt{pqt};p,q,t^2)},
\label{normell}\end{equation}
where
$$
\Gamma(z;p,q,t) \ = \  \prod_{j,k,l=0}^{\infty}
(1-zt^jp^kq^l)(1-z^{-1}t^{j+1}p^{k+1}q^{l+1}), \quad |p|,|q|,|t|<1,
$$
is the  second order elliptic gamma function
satisfying the  $t$-difference equation
$$
\Gamma(tz;p,q,t) \ = \ \Gamma(z;p,q) \Gamma(z;p,q,t).
$$
The reflection equation $\Gamma(z^{-1};p,q,t)=\Gamma(pqtz;p,q,t)$
leads to the equality
$$
M(x^{-1};p,q,t)M(x;p,q,t)=1.
$$
It is easy also to check validity of the functional equation
$$
M(x;p,q,t)M(t^{-1}x;p,q,t)=\Gamma\Big(x\sqrt{\frac{pq}{t}};p,q\Big),
$$
which is equivalent to \eqref{normeq} after fixing $t=pq$ and $x=e^{2\alpha}$.
Therefore we find the needed normalizing function
\begin{equation}
m(\alpha)=M(e^{2\alpha};p,q,pq), \quad m(\alpha)m(-\alpha)=1.
\label{ma1}\end{equation}
The function $-\log m(\alpha)$ defines thus the free energy per edge of
the integrable lattice model under consideration.

Now we discuss the partition function for the general
hyperbolic beta integral
solution of the star-triangle relation \eqref{modSTR}. The needed normalization
constant $m(\alpha)$ is found from the equation
\be
\frac{m(\alpha)}{m(\eta-\alpha)}\gamma^{(2)}(-2\alpha;\mathbf{\omega})=1,
\quad \text{or}\quad m(\alpha/2+\eta)=\gamma^{(2)}(\alpha;\mathbf{\omega})
m(-\alpha/2),
\label{modeqm}\ee
where $\eta=-(\omega_1+\omega_2)/2$.

Let us define the function
$$
\mu(u;\m_1,\m_2,\m_3)=
\frac{\gamma^{(3)}(u+\frac{1}{2}\sum_{k=1}^3\m_k+\m_3;\m_1,\m_2,2\m_3)}
{\gamma^{(3)}(u+\frac{1}{2}\sum_{k=1}^3\m_k;\m_1,\m_2,2\m_3)},
$$
where $\gamma^{(3)}$-function is the hyperbolic gamma function
of the third order defined in Appendix B.
Using the integral representation for it, we can write
\begin{equation}
 \mu(u;\mathbf{\omega})=
\exp\Big(- \frac{\pi \textup{i}a}{6}
-\int_{\R+\textup{i}0}\frac{e^{vx}}{(e^{\m_1 x}-1)(e^{\m_2 x}-1)(e^{\m_3 x}+1)}
\frac{dx}{x}\Big),
\label{mu}\end{equation}
where $v=u+\sum_{k=1}^3\m_k/2$ and
\begin{eqnarray*}&&
a=B_{3,3}(v+\omega_3;\m_1,\m_2,2\m_3)-B_{3,3}(v;\m_1,\m_2,2\m_3)
\\  && \makebox[0.5em]{}
= \frac{3}{2\m_1\m_2}\left(u^2-\frac{\m_1^2+\m_2^2+3\m_3^2}{12} \right)
\end{eqnarray*}
For a special choice of the third quasiperiod variable $\m_3=\m_1+\m_2$,
this function appeared for the first time in \cite{LZ}.

Using the reflection equation
$$
\gamma^{(3)}(\sum_{k=1}^3\m_k-u;\m_1,\m_2,\m_3)=
\gamma^{(3)}(u;\m_1,\m_2,\m_3)
$$
and the difference equation
$$
\gamma^{(3)}(u+\m_3;\m_1,\m_2,\m_3)=\gamma^{(2)}(u;\m_1,\m_2)
\gamma^{(3)}(u;\m_1,\m_2,\m_3),
$$
one can easily check that $\mu(u;\mathbf{\omega})\mu(-u;\mathbf{\omega})=1$ and
$$
\mu(u;\mathbf{\omega})\mu(u-\omega_3;\mathbf{\omega})=
\gamma^{(2)}(u+\frac{1}{2}(\m_1+\m_2-\m_3);\m_1,\m_2).
$$
The latter relation coincides with equation \eqref{modeqm}
for $u=2\alpha$ and $\m_3=\m_1+\m_2$.
Therefore we find the free energy per edge as $-\log m(\alpha)$, where
\begin{equation}
m(\alpha)=\mu(2\alpha;\m_1,\m_2,\m_1+\m_2).
\label{ma2}\end{equation}
By construction this function satisfies also the reflection equation $m(\alpha)m(-\alpha)=1$.
Denoting $\m_1=b, \m_2=b^{-1}$ and substituting the infinite product representation
of the $\gamma^{(3)}$-function given in Appendix B, we find the expression
\begin{eqnarray}\nonumber &&
m(\alpha)=\exp\left(-\pi\textup{i}\alpha^2-\frac{\pi\textup{i}}{24}(1-2(b+b^{-1})^2)\right)
\\ && \makebox[2em]{} \times
\frac{({\tilde q}e^{2\pi\textup{i}u/b};\tilde q^2)_\infty}
{(qe^{2\pi\textup{i}ub};q^2)_\infty}
\prod_{j,k=0}^\infty \frac{1+e^{\pi\textup{i}u/(b+b^{-1})}\tilde p^{j+1}\tilde q^{2k}}
{1-e^{\pi\textup{i}u/(b+b^{-1})}\tilde p^{j+1}\tilde q^{2k}},
\label{my}\end{eqnarray}
where it is assumed that $|q|<1$, $q=e^{2\pi \textup{i}b^2}$,
$\tilde q=e^{-2\pi \textup{i}/b^2}$,
and $\tilde p=e^{-\pi\textup{i}/(1+b^2)}$.

We turn now to the Faddeev-Volkov model solution of STR \eqref{astrFV}.
In this case we have no self-interaction of the spins sitting in lattice
vertices, and  the Boltzmann weights attached to edges are simplified. But the
partition function asymptotics is the same as in the previous case, since
evidently the normalizing constant $m(\alpha)$ is found from the same
equation \eqref{modeqm}.
The free energy per edge for this model was computed already
by Bazhanov, Mangazeev, and Sergeev in \cite{BMS}, where the Boltzmann
weights normalizing factor was denoted as $F_\theta$.
Comparing this constant with our $m(\alpha)$, we see that
they coincide for $\alpha=-(b+b^{-1})\theta/(2\pi)$, as necessary.
However, our infinite product representation of $m(\alpha)$  in \eqref{my}
differs drastically from that given in \cite{BMS} (which was the source of
author's old time confusion).

\section{Conclusion}

After the discovery of elliptic hypergeometric integrals,  for a long time the
author was drawing  attention of experts (including the second author of \cite{BS})
in two-dimensional conformal field theory and solvable models of statistical mechanics
for a potential emergence of such functions in these fields. The connection
between the elliptic beta integral and the star-triangle relation found in \cite{BS}
and the star-star relation described above confirms this expectation.
However, the nature appeared to be much richer than it was imagined in
\cite{spi:special,spi:talk,spi:essays}.
As mentioned already, the Dolan-Osborn discovery of a stunningly unexpected coincidence
of elliptic hypergeometric integrals with superconformal (topological) indices
in four dimensional supersymmetric gauge theories strongly pushed forward the
development of the theory and raised many  interesting open questions
\cite{DO,SV}. The interpretation of exact computability of the
elliptic beta integrals as the confinement phenomenon in quantum field theory
is a new type of conceptual perception of the exact mathematical
formulas.

As to the models considered in this paper,  we have
described a generalization of the Faddeev-Volkov solution of STR \cite{VF} with the
continuous spin variables taking values on the real line, which was not considered
in \cite{BS}.  It has some nontrivial self-interaction
energy for each vertex and a more complicated form of the Boltzmann weights
for edges, though the free energy per edge appears to be the same
as in the Faddeev-Volkov model.
In \cite{VF} the Yang-Baxter equation was proved using the quantum pentagonal
relation \cite{fkv}. It would be interesting to interpret in a similar way the model
we have described here. Some time ago the author has tried to find an
elliptic analogue of the pentagon relation in analogy with the constructions
described in \cite{volk}, but could not do it yet.
Clearly the elliptic beta integral gives already an analytic form of
that wanted operator relation, but it is hard to formulate it in
terms of the commutation relations of some explicit operators.

In \cite{FV}, Faddeev and Volkov have considered a lattice Virasoro algebra
and described an integrable model in the discrete $2d$
space-time (it was discussed also in detail in \cite{fkv}).
The elliptic beta integral yields more
general  solutions   of STR than that of \cite{VF}, and it is natural
to ask for explicit realization of the corresponding models
similar to \cite{FV}. During the work on \cite{SV},
G. Vartanov and the author have suggested that there should
exist some elliptic deformation of the primary fields $V_\alpha(z)$
built from free $2d$ bosonic fields
(in the spirit similar to the situation discussed in \cite{SWyl}) such
that the three point correlation function would be given by the elliptic
beta integral and the four point function would be described by the $V$-function
satisfying the elliptic hypergeometric equation \cite{spi:essays}
(so that the tetrahedral symmetries of the $V$-function would
describe the $s$-$t$ channels duality).
Unfortunately, such hypothetical results are not conceivable
at the present moment.

From the point of view of superconformal indices
the partition function associated with the elliptic beta integral solution of STR
looks as a superconformal index for a particular $SU(2)$-quiver gauge theory
on a two dimensional lattice. Recently there was a great deal of activity
on interrelations between four-dimensional super-Yang-Mills theories
and two-dimensional field theories,
see, e.g., \cite{alday,CNV,gprr,NS, SWyl}.
In this framework, the elliptic hypergeometric integrals
describing superconformal indices of $\mathcal{N}=2$
quiver gauge theories have been interpreted by Gadde et al
in \cite{gprr} as correlation functions of some $2d$
topological quantum field theories.

Therefore it is natural to expect that  superconformal
indices of all four dimensionanl CFTs
are related to discretizations of $2d$ CFT models
and other integrable systems. A connection of the Yang-Baxter moves
with the Seiberg duality has been briefly discussed in \cite{HV}.
In this context, superconformal indices of all quiver gauge theories
should correspond to full partition functions of some spin systems.
In view of the abundance of supersymmetric dualities and rich structure
of the corresponding superconformal indices (twisted partition functions) \cite{SV},
the author considers the present moment only as a beginning of uncovering new
two-dimensional and higher-dimensional integrable models hidden behind the
elliptic hypergeometric functions.

For instance, the elliptic Selberg integral defined on the $BC_n$
root system reads  \cite{die-spi:selberg,spi:essays}:
\begin{eqnarray}\nonumber
\kappa_n \int_{\T^n} \prod_{1\leq j<k\leq n}
\frac{\Gamma(tz_j^{\pm 1} z_k^{\pm 1};p,q)}{\Gamma(z_j^{\pm 1} z_k^{\pm 1};p,q)}
\prod_{j=1}^n\frac{\prod_{m=1}^6\Gamma(t_mz_j^{\pm 1};p,q)}{\Gamma(z_j^{\pm2};p,q)}
\frac{dz_j}{\textup{i}z_j}
\\
= \prod_{j=1}^n\left(\frac{\Gamma(t^j;p,q)}{\Gamma(t;p,q)}
\prod_{1\leq m<s\leq 6}\Gamma(t^{j-1}t_mt_s;p,q )\right),
\label{e-Selberg}\end{eqnarray}
where $|t|, |t_m|<1$, $t^{2n-2}\prod_{m=1}^6t_m=pq$, and
$$
\kappa_n =\frac{(p;p)_\infty^n (q;q)_\infty^n}{(4\pi)^n n!}.
$$
After some work, this formula can be given the STR type shape
\begin{eqnarray}\nonumber
&& \int_{[0,2\pi]^n} S({\bf u};t,p,q)W(\eta-\alpha;x,{\bf u})
W(\alpha+\gamma;y,{\bf u})W(\eta-\gamma;w,{\bf u})[d{\bf u}],
\\ && \makebox[2em]{}
=W_t(\alpha;y,w)W_t(\eta-\alpha-\gamma;w,x)
W_t(\gamma;x,y),
\label{astr_n}\end{eqnarray}
where we denoted
$$
[d{\bf u}]=\kappa_n \prod_{j=1}^n \frac{\Gamma(t;p,q)du_j}{\Gamma(t^j;p,q)},
$$
and the crossing parameter $\eta$ is defined as
$$
e^{-2\eta}=pqt^{n-1}.
$$
The Boltzmann weights have the form
\begin{eqnarray}
&& S({\bf u};t,p,q)=\prod_{1\leq j<k\leq n}
\frac{\Gamma(te^{\pm \textup{i} u_j \pm \textup{i} u_k};p,q)}
{\Gamma(e^{\pm \textup{i} u_j \pm \textup{i} u_k};p,q)}
\prod_{j=1}^n \frac{1}{\Gamma(e^{\pm 2\textup{i} u_j};p,q)}
\label{norms_n} \end{eqnarray}
and
\begin{eqnarray}\label{wbcn}
&& W(\alpha;x,{\bf u}):=\frac{1}{m(\alpha)}
\prod_{j=1}^n\Gamma(\sqrt{pq}e^{\alpha}e^{ \pm\textup{i}  x}
e^{\pm\textup{i} u_j};p,q),
\\
&& W_t(\alpha;x,y):=\frac{1}{m(\alpha)}
\prod_{j=1}^n\Gamma(\sqrt{pq}e^{\alpha}t^{j-\frac{n+1}{2}}e^{\pm \textup{i}x}
e^{\pm \textup{i} y};p,q),
\label{weight_n}\end{eqnarray}
and satisfy the reflection relations
$$
W(\alpha;x,{\bf u})W(-\alpha;x,{\bf u})=1, \qquad
W_t(\alpha;x,y)W_t(-\alpha;x,y)=1.
$$
The normalization constant $m(\alpha)$ for $n>1$ has a substantially more complicated
form than that for $n=1$. To describe it we introduce the
function
$$
M(x;p,q,t,s)=\frac{\Gamma(xts^2,xt^{1-n}s;p,q,t,s^2)}
{\Gamma(xts,xt^{1-n}s^2;p,q,t,s^2)},
$$
a particular ratio of four elliptic gamma functions of the {\em third order}.
More precisely, one has
$$
\Gamma(z;p,q,t,s):=\prod_{i,j,k,l=0}^\infty
\frac{1-z^{-1}p^{i+1}q^{j+1}t^{k+1}s^{l+1}}{1-zp^{i}q^{j}t^{k}s^{l}}
$$
for $z\in\C^*, \, |p|,|q|,|t|,|s|<1$, with the reflection
equation $\Gamma(z,pqtsz^{-1};p,q,t,s)=1$
and the difference equation $\Gamma(sz;p,q,t,s)=\Gamma(z;p,q,t)\Gamma(z;p,q,t,s)$.
Then,
\begin{equation}
m(\alpha)=M(e^{2\alpha};p,q,t,pqt^{n-1}),
\label{ma3}\end{equation}
with the standard reflection relation $m(\alpha)m(-\alpha)=1$.

Let us discuss a physical meaning of the obtained model. Consider a honeycomb
lattice on the plane with two types of vertices -- black and white with two adjacent
vertices always being of different color,
see the left-hand side of figure 2. Into each white vertex we put
an independent single component continuous spin $x$. Into each black
vertex we put $n$ independent spins $u_j,\, j=1,\ldots,n$, or one $n$-dimensional
spin with $n$ continuous components. These ``spins" are quite different from
those of the Ising model where they take
only the values $+1$ and $-1$ (i.e., they represent the fields
and not the compact spins). In different
words, one associates to each black vertex the $SP(2n)$-group space
related to the root system $BC_n$.
We associate with each black vertex the self-interaction Boltzmann weight
\eqref{norms_n} with an additional interaction between ``spin" components
in the internal space. To each bond connecting ``black-and-white" vertices
we attach the Boltzmann weight \eqref{wbcn}. Then, on the left-hand side
of \eqref{astr_n} we have the partition function of an elementary cell
with the black vertex in the center and the integral taken over the $u_j$-spin values.
If we apply this star-triangle relation to each black vertex we come
to a different spin system associated with the plain triangular lattice
having only the white vertices with the bond Boltzmann weights described
by the function \eqref{weight_n}, see the right-hand side of figure 2.
Such a transformation of lattices looks quite similar
to a transformation of the honeycomb-triangular Ising systems
considered in \cite{W}. Perhaps there exists also another STR type duality
transformation involving only the white vertices (with some self-interaction)
allowing for a transition to yet another triangular lattice system.
In addition to this uncertainty, it remains also unclear the free
energy per edge of which model is described by the function \eqref{ma3}.

Positivity of the Boltzmann weights of this model can be analyzed along
the lines of elliptic modular double involutions discussed in  \cite{spi:con}.
In particular, these weights are clearly positive
for $x, u_j\in [0,2\pi]$, real $t, \alpha$ and $\rho:=|\sqrt{pq}e^\alpha|<1$,
$\rho< |t|^{\frac{n-1}{2}}<\rho^{-1}$, with either $p^*=p, q^*=q$
or $p^*=q.$
For $n>1$ the crossing parameter $\eta$ looks like an arbitrary
free variable, not related to
other parameters of the system, but, in fact, it is essentially
equivalent to the coupling constant $t$ for $u_j$-spins.
If $t=1$, relation \eqref{astr_n} reduces to $n$-th power of the standard STR.

Define the $BC_n$-root system generalization of the $V$-function:
\begin{eqnarray}\label{type2} &&
I(t_1,\ldots,t_8;t;p,q)=\prod_{1\leq j<k\leq 8}\Gamma(t_jt_k;p,q,t)
\\ && \makebox[1em]{} \times
\kappa_n \int_{\T^n}\prod_{1\le j<k\le n}\!
\frac{\Gamma(tz_j^{\pm 1}z_k^{\pm1};p,q)}
{\Gamma(z_j^{\pm1}z_k^{\pm1};p,q)}
\prod_{j=1}^n \frac{\prod_{k=1}^8\Gamma(t_kz_j^{\pm1};p,q)}
{\Gamma(z_j^{\pm 2};p,q)}\frac{dz_j}{\textup{i} z_j},
\nonumber\end{eqnarray}
where parameters $t,t_1,\ldots,t_8\in\C$, satisfy
 $|t|, |t_j|<1$, and $ t^{2n-2}\prod_{j=1}^8t_j=p^2q^2$ constraints.
As shown by Rains \cite{Rains},
this function obeys the same $W(E_7)$ Weyl group of symmetries
as in the $n=1$ case. The key transformation has the form
\begin{equation} \label{ss_n}
I(t_1,\ldots,t_8;t;p,q)=I(s_1,\ldots,s_8;t;p,q),
\end{equation}
where
\begin{eqnarray*}
\left\{
\begin{array}{cl}
s_j =\rho^{-1} t_j,&   j=1,2,3,4  \\
s_j = \rho t_j, &    j=5,6,7,8
\end{array}
\right.;
\quad \rho=\sqrt{\frac{t_1t_2t_3t_4}{pqt^{1-n}}}
=\sqrt{\frac{pqt^{1-n}}{t_5t_6t_7t_8}},
\quad |t|,|t_j|,|s_j|<1.
\end{eqnarray*}
Introduce variables $x_j$ by relation
$t^{(n-1)/4}t_j=(pq)^{1/4}e^{2\pi\textup{i}x_j}$, so that the balancing
condition becomes $\sum_{j=1}^8x_j=0$. Then \eqref{ss_n} describes the
invariance of the integral $I$ with respect to the Weyl reflection
$$
x\to S_v(x)=x -\frac{2\langle x,v\rangle}{\langle v,v\rangle} v,
\quad x,v\in\mathbb{R}^8,
$$
where $\langle x,v\rangle=\sum_{k=1}^8x_kv_k$ is the scalar product
and the vector $v$ has components $v_k=1/2,\, k=1,2,3,4,$ and
$v_k=-1/2,\, k=5,6,7,8.$ Together with the group $S_8$ permuting
the parameters $x_j$, this transformation generates full exceptional
reflection group $W(E_7)$.

Equality \eqref{ss_n} can be rewritten in the star-star relation form
\begin{eqnarray}\nonumber && \makebox[-1em]{}
\int_{[0,2\pi]^n} S({\bf u};t,p,q)W(\eta-\alpha;x,{\bf u})
W(\eta-\beta;y,{\bf u})W(\gamma;w,{\bf u})W(\delta;z,{\bf u})[d{\bf u}]
\\ && \makebox[2em]{}
=R\frac{P_t(\alpha+\beta;x,y)}{P_t(\alpha+\beta;w,z)}
\int_{[0,2\pi]^n} S({\bf u};t,p,q)
\label{ss_BCn}\\ && \makebox[4em]{} \times
W(\beta;x,{\bf u})
W(\alpha;y,{\bf u})W(\eta-\delta;w,{\bf u})W(\eta-\gamma;z,{\bf u})[d{\bf u}],
\nonumber \end{eqnarray}
where $\alpha+\beta=\gamma+\delta$ and
$$
R=\prod_{l=0}^{n-1}\frac{\Gamma(t^{-l}e^{-2\alpha},t^{-l}e^{-2\beta};p,q)}
{\Gamma(t^{-l}e^{-2\gamma},t^{-l}e^{-2\delta};p,q)}, \quad
P_t(\alpha;x,y)=\prod_{l=0}^{n-1}\Gamma(t^{-l}e^{-\alpha}
 e^{\pm\textup{i}x\pm\textup{i}y};p,q).
$$
In complete analogy with $n=1$ case \eqref{t1}, \eqref{t2}, \eqref{t3},
one can obtain two other
differently looking star-star relations for $n>1$ by an iterative
application of this formula after permutations of parameters.

Using the matrix integral representations for elliptic hypergeometric integrals,
in \cite{SV} relation \eqref{ss_n} was shown to describe a new electric-magnetic
duality between two four-dimensional
$\mathcal{N}=1$ supersymmetric Yang-Mills theories with the gauge group
$G=SP(2n)$.
Namely, the electric theory has the flavor group $SU(8)\times U(1)$;
it contains the vector superfield in the adjoint
representation of $G$, one chiral scalar multiplet in the
fundamental representations of $G$ and $SU(8)$, and the field
described by the antisymmetric tensor of the second rank of $G$.
The magnetic theory has the flavor group
$SU(4)_l\times SU(4)_r\times U(1)_B\times U(1)$
and a similar set of quantum fields, as well as $2n$ additional gauge invariant
mesonic fields --- the antisymmetric tensors of $SU(4)$-flavor subgroups.
The elliptic Selberg integral \eqref{e-Selberg} describes the confinement
phenomenon in the $SP(2n)$ super-Yang-Mills gauge theory with
$6$  chiral superfields in the fundamental and 1 chiral superfield
in the antisymmetric representations of $SP(2n)$, respectively, ---
its dual magnetic phase contains only a peculiar set of mesonic
fields without local gauge symmetry.

From the present paper point of view
relations \eqref{astr_n} and \eqref{ss_BCn}
should have an appropriate physical interpretation in the
context of discrete integrable models for $n>1$ similar to the $n=1$ case.
We described already one possible honeycomb lattice model that can be associated with
the elliptic Selberg integral. The system lying behind relation \eqref{ss_BCn}
resembles the checkerboard Ising model with the continuous spins.
As an elementary cell one has a square with four white vertices
(with the single component spins $x,y,\ldots$ sitting in them) and one black vertex
in the center (with the $n$-component spin $\bf u$ sitting in it
and the integral taken over its values), see figure 4.
There are again three differently looking star-star
relations for $n>1$ obtained by repeated application of the
same formula \eqref{ss_BCn} in conjugation with permutation of parameters,
quite similar to the $n=1$ case. Equality \eqref{astr_n} can be considered
then as their reduction to STR.

We did not discuss in this paper an important
physical question of the existence of phase transitions in the described
models and the spectrum of scaling exponents. For clarifying this point
it is necessary to single out the temperature like variable associated
with one of the parameters $p$ or $q$ \cite{BS} and investigate the
behavior of the partition functions per edge (defined by $m(\alpha)$'s)
when the temperature varies from large to small values.
Since many known systems with nontrivial phase transitions are represented
by the limiting cases of the elliptic beta integral STR solutions, there
are nontrivial critical phenomena. However, their classification
requires separate analysis and lies beyond the scope of the present work.

In \cite{SV} a large number of proven and conjectural evaluation formulas
for elliptic beta integrals on root systems and their nontrivial symmetry transformation
 analogues for higher order integrals has been listed.
Actually, it was conjectured that there exist infinitely many such integrals,
and for each of them one can expect suitable application in
the context of solvable models of statistical mechanics and
other types of integrable systems.

\smallskip
{\bf Acknowledgments.}
The results of this paper were partially reported at the Jairo Charris
seminar (3-6 August 2010, Santa Marta, Colombia). The hospitality of P. Acosta-Humanez
during this workshop is gratefully appreciated. The author is deeply indebted
to A.N. Kirillov, V.B. Priezzhev, I.P. Rochev, and G.S. Vartanov for
stimulating discussions. A.M. Povolotsky is thanked for teaching me
the graphics drawing. Both referees are thanked for helping in improving
the paper.

\appendix
\section{The modified $q$-gamma function}

The function $\gamma(u;\omega_1,\omega_2)$ \eqref{2d-sin}
or its various transformed versions are
referred to in different papers as the double sine function \cite{kls,PT},
the non-compact quantum dilogarithm \cite{fad:mod,fkv,PT,volk,BS},
the hyperbolic gamma function \cite{rui,bult}, or the modified $q$-gamma
function \cite{spi:essays}.

The functional equations satisfied by $\gamma(u;\mathbf{\omega})$ have the form
\begin{equation}
\frac{\gamma(u+\omega_1;\mathbf{\omega})}{\gamma(u;\mathbf{\omega})}=
 1-e^{2\pi \textup{i}  \frac{u}{\omega_2}},\qquad
\frac{\gamma(u+\omega_2;\mathbf{\omega})}{\gamma(u;\mathbf{\omega})}=
 1-e^{2\pi \textup{i}  \frac{u}{\omega_1}}.
\label{func-eq}\end{equation}
Using a modular transformation for theta functions one can derive
another representation for $\gamma(u;\omega_1,\omega_2)$
complementary to \eqref{2d-sin}:
\begin{equation}
\gamma(u;\omega_1,\omega_2) = e^{\pi \textup{i} B_{2,2}(u;\mathbf{\omega})}
\frac {(e^{-2\pi \textup{i}  u/\omega_2}q; q)_\infty}
{(e^{-2\pi \textup{i} u/\omega_1}; \tilde q)_\infty}.
\label{modtr}\end{equation}

The non-compact quantum dilogarithm \cite{fad:mod}
in the notation of \cite{BMS} (in \cite{fkv} it was denoted as $e_b(z)$)
has the form
\begin{eqnarray*} &&
\varphi(z)=\exp\left(\frac{1}{4}\int_{\R+\textup{i} 0}\frac{e^{-2izw}}{\sinh(wb)\sinh(w/b)}
\right)\frac{dw}{w}
\\ && \makebox[2em]{}
=\exp\left(\int_{\R+\textup{i} 0}
\frac{e^{wu}}{(e^{wb}-1)(e^{w/b}-1)}
\right)\frac{dw}{w},
\end{eqnarray*}
where
$$
u=\frac{1}{2}(b+b^{-1})-\textup{i}z.
$$
For $q=e^{2\pi \textup{i}  b^2}$, $\tilde q=e^{-2\pi \textup{i} / b^2}$
and Im$(b^2)>0$, one can write
$$
\varphi(z)=\frac{(e^{2\pi \textup{i}  bu}; q)_\infty}
{(e^{2\pi \textup{i} u/b}\tilde q; \tilde q)_\infty}
=\frac{(-q^{1/2}e^{2\pi bz}; q)_\infty}
{(-\tilde q^{1/2} e^{2\pi z/b}; \tilde q)_\infty}.
$$
Therefore,
$$
\varphi(z)=\gamma \big(\frac12 (b+b^{-1})-\textup{i}z;b,b^{-1}\big)^{-1}.
$$
The $S_b(u)$ function used in \cite{PT} coincides with
$$
\gamma^{(2)}(u;\omega_1,\omega_2)=e^{-\frac{\pi \textup{i}}{2}B_{2,2}(u;\omega_1,\omega_2)}
\gamma(u;\omega_1,\omega_2)
$$
for $\omega_1=b$ and $\omega_2=b^{-1}$, and another similar function
of  \cite{PT} is
$$
G_b(u)=e^{-\frac{\pi \textup{i}}{12}(3+b^2+b^{-2})}\gamma(u;b,b^{-1}).
$$
The $\gamma(z)$-function used in \cite{volk}
coincides with the infinite products ratio on the right-hand side of  \eqref{modtr}
for $z=\omega_2u$ and $\tau=\omega_1/\omega_2$.

For $\text{Re}(\omega_1), \text{Re}(\omega_2)>0$ and
$0<\text{Re}(u)<\text{Re}(\omega_1+\omega_2)$ the function
$\gamma^{(2)}(u;\omega_1,\omega_2)$ has the following
integral representation
\be
\gamma^{(2)}(u;\omega_1,\omega_2)=\exp\left(-\text{PV}\int_\R
\frac{e^{ux}}
{(1-e^{\omega_1 x})(1-e^{\omega_2 x})}\frac{dx}{x}\right),
\label{bargamma}\ee
where the principal value of the integral means
$\text{PV}\int_\R=2^{-1}(\int_{\R+\textup{i} 0}+\int_{\R-\textup{i} 0})$.
Using the fact that $\text{PV}\int_\R dx/x^k=0$ for $k>1$, one can write
$$
\gamma^{(2)}(u;\omega_1,\omega_2)=\exp\left(-\int_0^\infty
\Big(\frac{\sinh (2u-\omega_1-\omega_2)x}
{2\sinh (\omega_1 x)\sinh (\omega_2 x)} - \frac{2u-\omega_1-\omega_2}{2\omega_1\omega_2x}
\Big)\right)\frac{dx}{x}.
$$
Comparing this expression with the hyperbolic gamma function
$G_h(z;\mathbf{\omega})$ defined in \cite{rui}, one can see that
$$
G_h(z;\mathbf{\omega})=\gamma^{(2)}
\Big(\frac{1}{2}(\omega_1+\omega_2)-\textup{i}z;\mathbf{\omega}\Big),
\quad \text{Re}(\omega_{1}), \text{Re}(\omega_{2})>0.
$$
 Changing in \eqref{bargamma} the sign of the
integration variable $x\to -x$ and simultaneously $\omega_k\to -\omega_k,\, u\to -u$,
we find that
\be
\gamma^{(2)}(u;\omega_1,\omega_2)=\exp\left(+\text{PV}\int_\R
\frac{e^{ux}}
{(1-e^{\omega_1 x})(1-e^{\omega_2 x})}\frac{dx}{x}\right),
\label{bargamma'}\ee
where $\text{Re}(\omega_1), \text{Re}(\omega_2)<0$ and
$\text{Re}(\omega_1+\omega_2)<\text{Re}(u)<0$.

The double sine function is defined as
$S_2(u;\mathbf{\omega})=1/\gamma^{(2)}(u;\mathbf{\omega})$
and its properties were described in detail in the Appendix
of \cite{kls}. For the $\gamma^{(2)}$-function we have
$$
\gamma^{(2)}(u;\omega_1,\omega_2)^*=\gamma^{(2)}(u^*;\omega_1^*,\omega_2^*),\quad
\gamma^{(2)}(\frac{\omega_1+\omega_2}{2}\pm u;\omega_1,\omega_2)=1,
$$
and $\gamma^{(2)}(au;a\omega_1,a\omega_2)=\gamma^{(2)}(u;\omega_1,\omega_2)$ for
arbitrary complex $a\neq 0$. After such a rescaling in \eqref{bargamma'}
with $a=2\pi\textup{i}$ one gets the definition of the hyperbolic
gamma function given in \cite{rai:limits}.

The asymptotics we are interested in for $\text{Im}(\omega_1/\omega_2)>0$
have the form
\begin{eqnarray}\nonumber
&&
\lim_{u\to \infty}e^{\frac{\pi \textup{i}}{2}B_{2,2}(u;\mathbf{\omega})}
\gamma^{(2)}(u;\mathbf{\omega})=1, \qquad \text{for}\ \arg \omega_1 <\arg u
<\arg \omega_2+\pi,
\\ &&
\lim_{u\to \infty}e^{-\frac{\pi \textup{i}}{2}B_{2,2}(u;\mathbf{\omega})}
\gamma^{(2)}(u;\mathbf{\omega})=1, \qquad \text{for}\ \arg \omega_1 -\pi<\arg u
<\arg \omega_2.
\nonumber
\end{eqnarray}

\section{General multiple gamma functions}

Barnes introduced a multiple zeta function as the following $m$-fold series \cite{bar}
$$
\zeta_m(s,u;\mathbf{\omega})=\sum_{n_1,\ldots,n_m=0}^\infty
\frac{1}{(u+\Omega)^s}, \qquad\Omega=n_1\omega_1+\ldots+n_m\omega_m,
$$
where $s, u\in\C$. This series converges for $\text{Re}(s)>m$ under the condition
that all $\m_j$ lie in one half-plane defined by a line passing through zero.
Because of the latter requirement,
the sequences $n_1\omega_1+\ldots+n_m\omega_m$ do not have
accumulation points on the finite plane for any $n_j\to+\infty$.
It is convenient to assume for definiteness that Re$(\m_j)>0$.

The function $\zeta_m(s,u;\mathbf{\omega})$ satisfies equations
\be
\zeta_m(s,u+\omega_j;\mathbf{\omega})-\zeta_m(s,u;\mathbf{\omega})
=-\zeta_{m-1}(s,u;\mathbf{\omega}(j)), \quad j=1,\ldots,m,
\lab{zeta-eq}\ee
where $\mathbf{\omega}(j)=(\omega_1,\ldots,\omega_{j-1},\omega_{j+1},\ldots,
\omega_m)$ and $\zeta_0(s,u;\mathbf{\omega})=u^{-s}$.
The Barnes multiple gamma function is defined by the equality
$$
\Gamma_m(u;\mathbf{\omega})
=\exp(\partial \zeta_m(s,u;\mathbf{\omega})/\partial s)\big|_{s=0}.
$$
It satisfies finite difference equations
\begin{equation}
\Gamma_m(u+\omega_j;\mathbf{\omega})
=\frac{1}{\Gamma_{m-1}(u;\mathbf{\omega}(j))}\, \Gamma_m(u;\mathbf{\omega}),
\qquad j=1,\ldots,m,
\label{bar-eq}\end{equation}
where  $\Gamma_0(u;\omega):=u^{-1}$.

The multiple sine-function is defined as
$$
S_m(u;\mathbf{\omega})=\frac{\Gamma_m(\sum_{k=1}^m\m_k-u;\mathbf{\omega})^{(-1)^m}}
{\Gamma_m( u;\mathbf{\omega})}.
$$
It is more convenient to work with the hyperbolic gamma function
$$
\gamma^{(m)}(u;\mathbf{\omega})=S_m(u;\mathbf{\omega})^{(-1)^{m-1}}
$$
satisfying the equations
$$
\gamma^{(m)}(u+\omega_j;\mathbf{\omega})
= \gamma^{(m-1)}(u;\mathbf{\omega}(j))\, \gamma^{(m)}(u;\mathbf{\omega}),
\qquad j=1,\ldots,m.
$$
Note that the elliptic gamma function can be written as
a special combination of four Barnes gamma functions
of the third order \cite{spi:essays}, and similar relations
are valid for higher order elliptic gamma functions used
in the present paper.

One can derive the integral representation \cite{nar}
\begin{eqnarray*} &&
\gamma^{(m)}(u;\mathbf{\omega})=
\exp\left( -\text{PV}\int_{\R} \frac{e^{ux}}
{\prod_{k=1}^m(e^{\omega_k x}-1)}\frac{dx}{x}\right)
\\ && \makebox[2em]{}
= \exp\left(- \frac{\pi \textup{i}}{m!}B_{m,m}(u;\mathbf{\omega})
-\int_{\R+\textup{i}0}\frac{e^{ux}}
{\prod_{k=1}^m(e^{\omega_k x}-1)}\frac{dx}{x}\right)
\\ && \makebox[2em]{}
= \exp\left(\frac{\pi \textup{i}}{m!}B_{m,m}(u;\mathbf{\omega})
-\int_{\R-\textup{i}0}\frac{e^{ux}}
{\prod_{k=1}^m(e^{\omega_k x}-1)}\frac{dx}{x}\right),
\end{eqnarray*}
where Re$(\omega_k)>0$ and $0<\text{Re}(u)< \text{Re}(\sum_{k=1}^m\omega_k)$
and $B_{m,m}$ are multiple Bernoulli polynomials defined by the generating function
\be
\frac{x^m e^{xu}}{\prod_{k=1}^m(e^{\m_k x}-1)}
=\sum_{n=0}^\infty B_{m,n}(u;\m_1,\ldots,\m_m)\frac{x^n}{n!}.
\lab{ber}\ee

Infinite product representations for these functions have been derived in
\cite{nar}. In particular, for $|p|, |q|<1$ and $|r|>1$ we have
$$
\gamma^{(3)}(u;\mathbf{\omega})=e^{-\frac{\pi\textup{i}}{6}B_{3,3}(u;\mathbf{\omega})}
\prod_{j,k=0}^\infty \frac{(1-e^{{2\pi\textup{i}u}/{\m_1}}\tilde q^{j+1}r^{-(k+1)})
(1-e^{{2\pi\textup{i}u}/{\m_2}}p^{j}q^{k})}
{1-e^{{2\pi\textup{i}u}/{\m_3}}\tilde p^{j+1}r^{-k}},
$$
which is used in the main text after the reduction $\m_3=2(\m_1+\m_2)$
(or $p=q^2,$ $r=\tilde q^{-2}$, $\tilde p=e^{-\pi\textup{i}\m_2/(\m_1+\m_2)}$).

The functions $m(\alpha)$ \eqref{ma1}, \eqref{ma2}, and \eqref{ma3}
defining the free energy per edge as described in the main part of the paper
are related to particular cases of
the Lerch type generalization of the Barnes zeta-function:
$$
\zeta_m(s,u;\beta;\mathbf{\omega})=\sum_{n_1,\ldots,n_m=0}^\infty
\frac{\prod_{k=1}^m \beta_k^{n_k}}{(u+\Omega)^s}, \qquad
\Omega=n_1\omega_1+\ldots+n_m\omega_m,
$$
converging for all $|\beta_k|<1$, or $\text{Re}(s)>m$ and $|\beta_k|=1$ (provided
the same constraints on $\m_j$ are valid as in the plain Barnes case).
The univariate case, i.e. the proper Lerch zeta-function, is described,
e.g., in \cite{whi-wat}.

The function $\zeta_m(s,u;\beta;\mathbf{\omega})$ satisfies the following set
of finite difference equations
\be
\beta_j\zeta_m(s,u+\omega_j;\beta;\mathbf{\omega})-\zeta_m(s,u;\beta;\mathbf{\omega})
=-\zeta_{m-1}(s,u;\beta(j);\mathbf{\omega}(j)), \quad j=1,\ldots,m,
\ee
where $\mathbf{\omega}(j)=(\omega_1,\ldots,\omega_{j-1},\omega_{j+1},\ldots,
\omega_m)$, $\beta(j)=(\beta_1,\ldots,\beta_{j-1},\beta_{j+1},\ldots,
\beta_m)$,   and $\zeta_0(s,u;\beta;\mathbf{\omega})=u^{-s}$.

Similar to the Barnes case, one can easily derive the integral representations
\begin{eqnarray*} &&
\zeta_m(s,u;\beta;\mathbf{\omega})=\frac{1}{\Gamma(s)}
\int_0^\infty \frac{t^{s-1}e^{-ut}}{\prod_{k=1}^m(1-\beta_ke^{-\omega_k t)} } dt
\\ && \makebox[2em]{}
=\frac{\textup{i}\Gamma(1-s)}{2\pi}
\int_{C_H} \frac{(-t)^{s-1}e^{-ut}}{\prod_{k=1}^m(1-\beta_ke^{-\omega_k t}) } dt,
\end{eqnarray*}
where $C_H$ is the Hankel contour encircling the half-line $[0,\infty)$
counterclockwise, and using them analytically continue $\zeta_m$-function in $s$
and $\beta_k$ to different regions of parameters.
The $\beta_k$-deformation of the Barnes multiple gamma function defined
as $\Gamma_m(u;\beta;\mathbf{\omega})
=\exp(\partial \zeta_m(s,u;\beta;\mathbf{\omega})/\partial s)\big|_{s=0}
$
satisfies the finite difference equations
\begin{equation}
\Gamma_m(u+\omega_j;\beta;\mathbf{\omega})^{\beta_j}
=\frac{1}{\Gamma_{m-1}(u;\beta(j);\mathbf{\omega}(j))}\,
\Gamma_m(u;\beta;\mathbf{\omega}),\qquad j=1,\ldots,m,
\label{bar-ler-eq}\end{equation}
where  $\Gamma_0(u;\beta;\omega):=u^{-1}$.

When $\beta_k$ are primitive roots of unity, $\beta_k^{n_k}=1$, $n_k=2,3,\ldots,$
it is possible to rewrite $\zeta_m(s,u;\beta;\mathbf{\omega})$ as
linear combinations of the standard Barnes zeta functions.
It follows from the simple identity
$$
\frac{1}{1-\beta_kz}=\frac{\prod_{l=0,2,\ldots,n_k-1} (1-\beta_k^lz) } {1-z^{n_k}}.
$$
This allows expressing the functions like \eqref{mu} as linear
combinations of the standard Barnes gamma functions, which was used in the
construction of infinite product representations of the functions $m(\alpha)$
\eqref{ma1}, \eqref{ma2}, and \eqref{ma3}.
In particular, function \eqref{mu} is emerging from the $m=3$ case
with the choice $\beta_1=\beta_2=1,\, \beta_3=-1$.

\bibliographystyle{amsalpha}

\end{document}